\documentstyle[aasms4]{article}
\def\simless{\mathbin{\lower 3pt\hbox
   {$\rlap{\raise 5pt\hbox{$\char'074$}}\mathchar"7218$}}} 
\def\simgreat{\mathbin{\lower 3pt\hbox
   {$\rlap{\raise 5pt\hbox{$\char'076$}}\mathchar"7218$}}} 
\def\delvec{{\mbox{\boldmath $\delta$}}}
\def\muvec{{\mbox{\boldmath $\mu$}}}
\def\omvec{{\mbox{\boldmath $\omega$}}}
\def\nuvec{{\mbox{\boldmath $\nu$}}}

\def\betap{\beta^\prime}
\def\zetap{\zeta^\prime}
\def\etap{\eta^\prime}
\def\rhos{\rho_s}
\def\vorthat{\nuvec}
\def\vsvec{{\bf v_s}}
\def\vlvec{{\bf v_L}}
\def\vnvec{{\bf v_n}}
\def\uvec{{\bf u}}
\def\curl{\nabla\times}
\def\vort{\omvec}
\def\Omvec{{\bf\Omega}}
\def\Omveca{\Omvec_a}
\def\Omvecatil{\Omveca^{(+)}}
\def\Omvecb{\Omvec_b}
\def\Omvecbtil{\Omvecb^{(+)}}
\def\Omsvec{\Omvec_s}
\def\Omnvec{\Omvec_n}
\def\Omcore{\Omvec_{c}}
\def\Omcr{\Omvec_{cr}}
\def\Omcrtil{\Omega_{cr}^{(+)}}
\def\Omcortil{\Omega_c^{(+)}}
\def\Omcrone{\Omega_{cr,1}}
\def\Omcrtwo{\Omega_{cr,2}}
\def\Omcrthree{\Omega_{cr,3}}
\def\Omcorone{\Omega_{c,1}}
\def\Omcortwo{\Omega_{c,2}}
\def\Omcorthree{\Omega_{c,3}}
\def\Lcore{{\bf L_c}}

\def\Lp{{\bf L_p}}

\def\Icr{{\bf I_{cr}}}
\def\fdrag{{\bf F_d}}
\def\be{\begin{equation}}
\def\ee{\end{equation}}
\def\Nvec{{\bf N}}
\def\Nveccr{\tilde N_{cr}^{(+)}}

\def\Nveccor{\tilde N_c^{(+)}}
\def\Nveccrthree{\tilde N_{cr,3}}

\def\nvec{{\bf N_\beta}}
\def\nvecp{{\bf N_{\beta^\prime}}}
\def\fvec{{\bf f}}
\def\rvec{{\bf r}}
\def\Tbeta{{\bf T}_\beta}
\def\Tr{{\rm Tr}}
\def\Tbetap{{\bf T}_{\beta^\prime}}
\def\Omsone{\Omega_{s,1}}
\def\Omstwo{\Omega_{s,2}}
\def\Omsthree{\Omega_{s,3}}
\def\sigmap{\sigma^\prime}
\def\sigmapp{\sigma^{\prime\prime}}
\def\sigmapxi{\sigmap_\xi}
\def\Omstil{\Omega_s^{(+)}}

\def\Delomthree{\Delta_3}
\def\Delomtil{\Delta^{(+)}}

\def\epsone{\epsilon_1}
\def\epstwo{\epsilon_2}
\def\phip{\phi^\prime}
\def\Delomcor{\Delta_c^{(+)}}
\def\Bvar{i(1-\betap)+\beta}
\def\Zvar{i(1-\zetap)+\zeta}
\def\BmZvar{\Bvar-i(1-\zetap)-\zeta}

\begin{document}
\bigskip
\title{PRECESSION OF ISOLATED NEUTRON STARS I: \\
EFFECTS OF IMPERFECT PINNING}
\author{Armen Sedrakian, Ira Wasserman and James M. Cordes}
\affil{Center for Radiophysics and Space Research, Cornell University, Ithaca, NY 14853}

\begin{abstract}
We consider the precession of isolated neutron stars in which superfluid
is not pinned to the stellar crust perfectly. In the case of perfect pinning,
Shaham (1977) showed that there are no slowly oscillatory, long-lived modes.
When the assumption of perfect pinning is relaxed, new modes are found that
can be long-lived, but are expected to be damped rather than oscillatory,
unless the drag force on moving superfluid vortex lines has a substantial
component perpendicular to the direction of relative motion.
The response of a neutron star to external torques, such as the spindown
torque, is also treated. We find that when computing the response of a star
to perturbations,
assuming perfect coupling of superfluid to normal matter from the start can
miss some effects.
\end{abstract}

\section{INTRODUCTION}\label{sec:intro}

Radiopulsars can be exceptionally stable clocks. The predictability
of pulse arrival times has made possible precision tests of 
General Relativity (e.g. Taylor et al. 1992) and the discovery
of the first extrasolar planetary systems (Wolszczan and Frail 1992). However,
imperfections in the rotation of most pulsars have been monitored
for some time, most notably glitches (e.g. Boynton et al. 1969, 
Radhakrishnan and Manchester 1969, Backus et al.
1982, Downs 1982, Demianski and Proszynski 1983, Manchester et al. 1983,
Lyne 1987, Lyne and Pritchard 1987, Cordes, Downs and Krause-Polstorff 1988,
McKenna and Lyne 1990, Hamilton et al. 1989,
McCulloch et al. 1990, Flanagan 1990, 1993, Lyne, Smith and
Pritchard 1992, Shemar and Lyne 1996) and timing noise 
(e.g. Boynton et al. 1972, Groth 1975, Cordes and Helfand 1980, Cordes and Downs
1985, D`Alessandro et al. 1995). These deviations 
from stable spin convey information about
the internal structure and dynamics of neutron stars. For example,
the long-time healing of the spin frequencies and spindown rates
of glitching pulsars may be explained if the convulsions are due
to sudden unpinning of superfluid vortex lines that were glued
to nuclei in the pulsar's crust before the glitch, migrate as a consequence
of the glitch, and repin to other nuclei in its aftermath
(e.g. Anderson and Itoh 1975, Alpar et al. 1981, 1984a,b, 1993, Link,
Epstein and Baym 1993).

A small number of neutron stars also exhibit long term cyclical
but not precisely oscillatory variations in their spin. A particularly
well-known case is the Crab Pulsar, whose 
phase residuals (after careful fitting that accounts
for spindown and glitches) vary systematically with a peak-to-peak
range of order $\pm 10$ ms, and a characteristic cycle duration of about
20 months. (Lyne, Pritchard and Smith 1988). After
its Christmas 1988 glitch, the Vela Pulsar showed -- after accounting
for exponential recovery from the glitch -- damped oscillatory 
phase residuals with a period of order 25 days; evidence for oscillations
in the frequency derivative of the pulsar both before and after
the glitch with a period of ``a few tens of days'' was also
reported (McCulloch et al. 1990). Evidence for long term variations (correlation
times $\sim 100$ days) in the pulse shape of the Vela Pulsar
has also been found in data spanning approximately four years
(Blaskiewicz 1992, Cordes 1993). A principal component analysis of
the pulse shape of PSR 1642-03 (Blaskiewicz 1992; Blaskiewicz and Cordes
1997, in preparation) yields evidence for cyclical pulse shape
variations with a period of about 1000 days; long-term variations
on a similar characteristic timescale are also seen in the timing
residuals for this pulsar (Cordes 1993).  Finally, although Her X-1
is an accreting X-ray pulsar, not an isolated radiopulsar, it has
a well-known 35 day cycle on which it appears and disappears;
observed variations in pulse shape over the cycle suggest that
it is related to periodic variations in the rotation of the neutron
star (e.g. Tr\"umper et al. 1986, Alpar and \"Ogelman 1987).

Soon after the discovery of radiopulsars, it was suggested that long
term variations in their spin could result from free precession (e.g.
 Davis and Goldstein 1970, Goldreich 1970, Ruderman 1970, Brecher
1972, Pines et al. 1973, Pines and Shaham 1972, 1974, Lamb et al. 1975,
Jones 1976),
and that pulsar arcanae such as drifting subpulses might be related
to precessional effects.
If a neutron star were a rigid or semirigid solid (see Pines and
Shaham 1972, 1974), it would precess with a period of order $P/\epsilon$, 
where $P$ is the spin period of the star and
$\epsilon$ is the fractional difference between its principal moments
of inertia; $\epsilon\simless 10^{-7}$ would imply precession periods
$\simgreat P({\rm s})$ years, where $P({\rm s})$ is the spin period
in seconds.
However, once it became apparent that superfluid vortex lines pin
to the crust of a neutron star, Shaham (1977) demonstrated that slow,
persistent precession is impossible. Pinned superfluid alters the
effective asymmetry of the star to $I_p/I$, where $I_p$ is the moment
of inertia of the pinned superfluid and $I$ the moment of inertia
of the star (or stellar crust, depending on various coupling 
parameters); since $I_p/I\sim 10^{-2}$ or even larger, precession is
very fast. Moreover, Shaham (1977) demonstrated that precession would
decay rapidly for some estimates of the coupling timescale 
between crust and core (e.g. Alpar and Sauls 1988;  Sedrakian and
Sedrakian 1995).

Perhaps because of Shaham's (1977) pessimistic conclusions, there has
been relatively little theoretical work on the long term variability
of pulsar spins. Some 
argue that vortex line pinning does not occur, making
slow precession possible (e.g. Jones 1988 for the
Crab Pulsar), but it may be hard to support the viewpoint that
pinning is completely absent in 
face of physical arguments to the contrary (e.g. Alpar et al. 1984a,
Epstein and Baym 1988, Link and Epstein 1991).
Notwithstanding the pessimism of theorists, the data
demand an explanation. Ruderman (1997) has mentioned long term variations
in pulsar spin rates as one of the outstanding unresolved problems
of neutron star physics. 

This paper is the first of two that attack 
this problem. Our purpose in this paper is three fold: First, we 
revisit the arguments put forward by Shaham (1977) with an eye 
toward identifying possible loopholes. Although it will become
apparent that there are several different possibilities, we 
concentrate here on Shaham's assumption that superfluid pins perfectly
to crustal nuclei. (Some of the other loopholes will be 
considered in a subsequent paper.) We develop the formalism for
doing so in Section \ref{sec:setup}, and solve the equations governing
the spin dynamics of a neutron star in succeeding sections to
varying degrees of complexity and realism. As expected, we do find
modes in addition to those found by Shaham (1977), but we also argue
that none of these modes is likely to be a long period, slowly damped
oscillation.
Second, we examine what can happen in a multicomponent star, in
which some regions contain pinned superfluid, and others unpinned
superfluid. One might think that if some parts of a star are
capable of slow oscillations of the spin -- either precession or
long period fluid modes, such as Tkachenko modes -- then there could
be an observable signature of these modes in the detectable pulsar
spin rate. However, we demonstrate that this situation is highly
unlikely, for even if such regions do exist inside actual pulsars,
persistent long period oscillations in those domains are only
possible if the coupling to the crust, where superfluid is 
pinned effectively, is very weak; but under such conditions, the
crust is almost unaffected by the slow oscillations, which
hardly manifest themselves in the crustal spin rate. Third, we
begin a general examination of the effects of external torques
-- such as the spindown torque -- on the spin dynamics. For this
purpose, we derive explicit expressions for the response of 
the various components of a neutron star to rather general 
time dependent torques. 
Our treatment of this problem shows that the limit of perfect 
coupling must be taken carefully when the response to external
torques is needed, because the additional modes that appear
when pinning is imperfect contribute to the response, and
cannot be ignored.

As will become apparent in the succeeding sections, we do not
believe that the long term cyclic variability detected in the
spins of some pulsars can be accounted for by free precession,
and that it is not likely to be due to forced precession either. However,
we do believe that this paper begins to elucidate the complexity
of the behavior of neutron star spin, and clarifies the conditions
that must be met for precession to occur, even if those conditions
are not likely to be realized.

\section{OVERVIEW}\label{sec:overvu}

\subsection{Pinned Superfluid Suppresses Precession (Shaham 1977)}\label{sec:sh77}

Shaham (1977) showed that pinned crustal superfluid dramatically
alters the physics of precession. Let us review his argument
briefly. Consider a three component neutron star that consists
of: (i) a rigid crust rotating at angular velocity $\Omcr$;
(ii) pinned crustal superfluid, whose angular momentum $\Lp$ is 
independent of time in the frame rotating with the crust;
and (iii) a core (super)fluid rotating at angular velocity
$\Omcore$. As seen in the inertial frame,  
\be
I_c{d\Omcore\over dt}+{d(\Icr\cdot\Omcr)\over dt}+\Omcr\times\Lp=0,
\label{eq:angmom1}
\ee
if there are no external torques,
where $I_c$ is the moment of inertia of the core fluid and
$\Icr$ is the moment of inertia tensor of the crust. We assume that
the moment of inertia {\it tensor} of the core fluid is always of
the form ($\delvec$ is the unit tensor)
\be
{\bf I_c}=(I_c-\Delta I_c)\delvec+\Delta I_c\biggl({3\hat\Omcore\hat\Omcore
-\delvec\over 2}\biggr),
\ee
so that its angular momentum $\Lcore={\bf I_c}\cdot\Omcore=I_c\Omcore$.
This amounts to assuming that the core fluid adjusts its shape instantaneously
to an oblate spheroid flattened along its direction of rotation. We shall
discuss this assumption more fully in a subsequent publication.

Introducing a dissipative torque that seeks to enforce corotation between
the crust and the core, we get the coupled equations
\be
{d(\Icr\cdot\Omcr)\over dt}+\Omcr\times\Lp=-K(\Omcr-\Omcore)
\ee
\be
I_c{d\Omcore\over dt}=K(\Omcr-\Omcore),
\ee
where $K$ is a constant. These equations have a rich set of fixed points
(where time derivatives vanish)
depending on the orientation of $\Lp$ relative to the principal axes of
$\Icr$. The full set of fixed points, and their possible observable
significance, will be discussed completely elsewhere; here we focus on
the particularly simple -- but far from general -- situation in which
$\Lp$ is along one of the principal axes of $\Icr$. In that circumstance,
the fixed point solution is $\Omcr=\Omcore=\Omvec$, with $\Omvec\parallel
\Lp$.

Perturbations about this fixed point are studied most easily in the
frame corotating with the crust, where $\Icr$ is independent of time.
For definiteness, let us suppose that the principal moments of inertia
of the crust are $I_1<I_2<I_3$ and, to parallel Shaham (1977) as closely
as possible, suppose that the fixed point corresponds to rotation about
the 3-axis at angular velocity $\Omvec={\bf\hat e_3}\Omega$. Then
the linearized equations are
\be
\dot\Omcrone+\biggl[\biggl({I_3-I_2\over I_1}\biggr)+{L_p\over I_1\Omega}
\biggr]\Omcrtwo=-{K\over I_1\Omega}(\Omcrone-\Omcorone)
\ee
\be
\dot\Omcrtwo-\biggl[\biggl({I_3-I_1\over I_2}\biggr)+{L_p\over I_2\Omega}
\biggr]\Omcrone=-{K\over I_2\Omega}(\Omcrtwo-\Omcortwo)
\ee
\be
\dot\Omcrthree=-{K\over I_3\Omega}(\Omcrthree-\Omcorthree)
\ee
\be
\dot\Omcorone+\Omcrtwo-\Omcortwo={K\over I_c\Omega}(\Omcrone-\Omcorone)
\ee
\be
\dot\Omcortwo+\Omcorone-\Omcrone={K\over I_c\Omega}(\Omcrtwo-\Omcortwo)
\ee
\be
\dot\Omcorthree={K\over I_c\Omega}(\Omcrthree-\Omcorthree).
\ee
In these equations, $\dot{\bf F}=\Omega^{-1}d^\star{\bf F}/dt$, where $d^\star{\bf F}/dt$
is the time derivative of any vector
$\bf F$ as seen in the frame rotating with the crust.
It is clear that the perturbations along the 3-axis decouple from those along
other axes, and decay exponentially with a characteristic rate
\be
(\Omega\tau)^{-1}={K\over\Omega}\biggl({1\over I_3}+{1\over I_c}\biggr);
\ee
this coupling time has been estimated by Alpar and Sauls (1988), according
to whom $\Omega\tau/2\pi\approx 400-10^4$ 
(and the relaxation of the electron distribution function is
due to the scattering  off the neutron vortex magnetization), 
and by Sedrakian and Sedrakian
(1995), who find that $\Omega\tau/2\pi$ is rather sensitive to the 
mass density $\rho$ and spans the range  $\Omega\tau/2\pi\sim (10^2-10^8) P$(s)
for $\rho \sim (1.6 -3)\times 10^{14}$ g cm$^{-3}$ 
(here the electron scattering 
is off the proton vortex clusters coupled to the neutron vortex lattice).
In the latter case 
the relaxation time spans a wide density range with a large gradient near the 
crust-core interface  because of an exponential 
dependence of the size of the cluster on the proton effective mass.
(See also Sedrakian et al 1995, Sedrakian \& Cordes 1998; for
the decay of precession the effective 
coupling rate $\gamma$ is a weighted average of 
the range found by Sedrakian
and Sedrakian 1995, implying an effective 
coupling time closer to the smallest values; we adopt  
$\Omega\tau/2\pi\sim 100P({\rm s})$ for numerical estimates below.)

The remaining four equations have normal modes proportional to $\exp(p\Omega t)
\equiv\exp(p\phi)$, where $\phi=\Omega t$ is pulse phase.
It is straightforward to solve for the modes of a triaxial star, but the basic
result can be derived under the assumption of axisymmetry, $I_2=I_1$.
(We have solved the corresponding triaxial problem, and there are no qualitatively
different modes for slowly rotating neutron stars.)
If we define 
\be
\sigma=\biggl({I_3-I_1\over I_1}\biggr)+{L_p\over I_1\Omega}\qquad{\rm and}\qquad
\gamma={K\over \Omega}\biggl({1\over I_1}+{1\over I_c}\biggr)
={I_3(I_1+I_c)\over \Omega\tau I_1(I_3+I_c)},
\label{eq:sigdef}
\ee
then
\be
\dot\Omcrone+\sigma\Omcrtwo=-{\gamma I_c\over I_1+I_c}(\Omcrone-\Omcorone)
\ee
\be
\dot\Omcrtwo-\sigma\Omcrone=-{\gamma I_c\over I_1+I_c}(\Omcrtwo-\Omcortwo)
\ee
\be
\dot\Omcorone+\Omcrtwo-\Omcortwo={\gamma I_1\over I_1+I_c}(\Omcrone-\Omcorone)
\ee
\be
\dot\Omcortwo+\Omcorone-\Omcrone={\gamma I_1\over I_1+I_c}(\Omcrtwo-\Omcortwo).
\ee
If $\gamma=0$, so the crust and core are uncoupled,
then there are modes with $p=\pm i\sigma$ which correspond to
independent precession of $\Omcr$ but at a frequency that is much larger than
the conventional Euler frequency for reasonable values of $L_p/I_1\Omega
\equiv I_p/I_1$ (where $I_p$ is the moment of inertia of pinned superfluid).
The remaining modes with $p=\pm i$
are an artifact of working in the frame that corotates
with the crust, and correspond to $\Omcore$ fixed in the inertial frame
of reference.

When $\gamma\neq 0$, the modes are damped, as was discussed by Bondi and
Gold (1954) in the context of the rotation of the Earth (without considering
pinned superfluid, of course!). The characteristic equation is fourth order
in $p$, but we expect the roots to come in complex conjugate pairs, so
we can reduce the characteristic equation to second order by introducing
the complex angular velocities
\be
\Omcrtil=\Omcrone+i\Omcrtwo\qquad{\rm and}\qquad
  \Omcortil=\Omcorone+i\Omcortwo,
\ee
which satisfy the equations
\be
\dot\Omcrtil-i\sigma\Omcrtil=-{\gamma I_c\over I_1+I_c}(\Omcrtil-\Omcortil)
\ee
\be
\dot\Omcortil-i\Omcortil+i\Omcrtil={\gamma I_1\over I_1+I_c}(\Omcrtil-\Omcortil).
\ee
Substituting $(\Omcrtil,\Omcortil)\propto\exp(p\phi)$ we find
\be
p^2+p[\gamma+i(1-\sigma)]
+\sigma\biggl(1-{i\gamma I_1\over I_1+I_c}\biggr)=0.
\ee
The normal modes of the fourth order system are the two solutions to this
quadratic equation, and their complex conjugates.

Although we can solve the second order characteristic equation exactly,
it is more instructive to find approximate solutions valid for small and
large crust-core coupling.
For small values of $\gamma$, we rewrite the characteristic equation as
\be
(p+i)(p-i\sigma)+\gamma\biggl(p
-{i\sigma I_1\over I_1+I_c}\biggr)=0;
\ee
this form separates terms of zeroth and first order in $\gamma$
explicitly. To first order in $\gamma$, the solutions are
\be
p_d=-i-{\gamma [1+\sigma I_1/(I_1+I_c)]\over 1+\sigma}
\label{eq:pdshwk}
\ee
and
\be
p_p=i\sigma-{\gamma\sigma I_c\over (I_1+I_c)(1+\sigma)}.
\label{eq:ppshwk}
\ee
For large values of $\gamma$, we rewrite the characteristic equation
as
\be
p-i\sigma{I_1\over I_1+I_c}
+\gamma^{-1}[p^2+ip(1-\sigma)+\sigma]=0;
\ee
this form is useful for expanding in powers of $\gamma^{-1}$.
In this case, the solutions to first order in $\gamma^{-1}$
are
\be
p_d=-\gamma
-i\biggr(1-{\sigma I_c\over I_c+I_1}\biggr)
\label{eq:pdshstr}
\ee
and
\be
p_p={i\sigma I_1\over I_1+I_c}
	 -{\sigma I_c\over \gamma (I_1+I_c)}
	 \biggl(1+{\sigma I_1\over I_1+I_c}\biggr).
\label{eq:ppshstr}
\ee
In each case, $p_d$ represents damping of the angular velocity
difference between crust and core, and $p_p$ is the precessing mode. 

For the coupling times estimated by, for example,
Alpar and Sauls (1988) or
Sedrakian and Sedrakian (1995),
the small $\gamma$ limit is the relevant one. Since $I_p/I_1
\gg I_3/I_1-1$, the precession period is far smaller than for 
a rigid body, approximately $I_1/I_p$ spin periods.
Moreover, the wobble damps away, lasting 
$\sim\gamma^{-1}$ precession periods: 
$\gamma^{-1}\approx 400-10^4$ according to Alpar and Sauls
(1988) and a reasonable estimate for the effective coupling is
$\gamma^{-1}\sim 100P({\rm s})$ for Sedrakian and Sedrakian (1995).
Even if $\gamma$ were large, the precession period
would be short, although it would be lengthened by a factor
$1+I_c/I_1$ relative to the small $\gamma$ limit, implying
a cycle $(I_c+I_1)/I_p$ spin periods long. The precession would
persist for approximately $I_1/I_c\Omega\tau$  precession
periods in this limit.
Since the crust-core coupling time must exceed the light travel time
across the star, $\tau>R/c\approx 0.03$ ms, and the damping
time for the precession must be $\simless 5000(I_1/I_c)P$ precession
periods, where
$P$ is the rotation period in seconds.

In neither limit is the precession either long period or 
persistent. From this pessimistic result, one concludes that 
free precession cannot account for the cyclical behavior seen
in long time monitoring of some pulsars. Moreover, to explain
the data, one must invoke an excitation mechanism that acts
relatively continuously, since it must fight the tendency for
neutron star wobbles to decay rapidly. The characteristic
cycle timescales of order months to years observed for these
pulsars must reflect the underlying processes responsible
for the continuous excitations.

\subsection{Imperfect Pinning}\label{sec:setup}

In demonstrating that persistent, long period precession is
impossible for neutron stars with pinned superfluid, Shaham
(1977) assumed perfect pinning. In actuality, superfluid
vortex lines will not pin to crustal nuclei absolutely. One
purpose of this paper is to see whether there are new
oscillatory modes that emerge when pinning is assumed to 
be strong but not perfect.

To study this problem, we adopt a somewhat idealized approach.
In actuality, the pinning of crustal superfluid  is a 
highly inhomogeneous process involving the interaction of
individual vortex lines and crustal nuclei. This coupling
is modelled by effective potentials highly localized around
discrete pinning sites in the vortex creep picture (e.g. 
Anderson and Itoh 1975, Alpar et al. 1984a, Link and Epstein
1991, Link et al. 1993)
and by scattering of particles by and Kelvon
excitation of moving vortex lines not pinned to crustal nuclei
(e.g. Epstein and Baym 1992, Jones 1991, 1992).
In our calculations, we use
smoothed hydrodynamical equations to describe the coupling
between the superfluid and normal components of the crust
macroscopically, using the formalism developed in
Khalatnikov (1965, Section 16). This formulation of the problem
is linked most naturally to a picture in which superfluid
vortex lines experience drag forces as they move through
a smooth medium of normal fluid, but also may be applied
directly in the vortex creep picture in the linear approximation
(i.e. when the difference between the angular velocities of the
superfluid and normal fluid are sufficiently small).

Shaham's results are recovered in the limit of perfect
coupling, that is, when the coefficients of mutual friction
are infinite. We can explore whether qualitatively new
modes appear when the mutual friction is strong, but
pinning is not perfect. As we shall see, no new slowly damped,
long period modes arise.

General formulae for the mutual friction force are given
in Khalatnikov (1965, Section 16). If the superfluid vorticity is defined to be
$\vort=\curl\vsvec$, where $\vsvec$ is the superfluid velocity,
then the net force per unit volume acting on the superfluid is
\be
{\bf f}=-\vort\times(\curl\lambda\vorthat)
-\betap\rhos\vort\times\uvec-\beta\rhos\vorthat\times(\vort\times\uvec)
+\gamma^\prime\rhos\vorthat\vort\cdot\uvec,
\ee
where $\rhos$ is the superfluid mass density, $\vorthat=\vort/\vert\vort\vert$,
and, for a normal fluid velocity $\vnvec$,
\be
\uvec\equiv\vnvec-\vsvec-{1\over\rhos}\curl\lambda\vorthat;
\ee
$\lambda=(\rhos\kappa/4\pi)\ln(d/\xi)$, where $\kappa$ is the quantum of
circulation per vortex line, $d$ is the effective intervortex separation,
and $\xi$ is the coherence length. The mutual friction force is defined to
be ${\bf f}+\vort\times(\curl\lambda\vorthat)/\rhos$.
The parameters $\beta$ and $\gamma^prime$
must be positive for the rate of energy dissipation resulting from 
$\bf f$ to be greater than zero locally.

Qualitatively, the terms in $\bf f$ involving $\vorthat$ arise from the
bending of vortex lines, and shall be neglected here. Of the remaining
contributions to $\bf f$, the two proportional to $\beta$ and $\betap$
are perpendicular to $\vort$ whereas the one proportional to $\gamma^\prime$
is along $\vort$; the latter is expected to be small, and we neglect
it too. With these simplifications, the form for $\bf f$ used in this
paper is
\be
{\bf f}=-\betap\rhos\vort\times(\vnvec-\vsvec)
-\beta\rhos\vorthat\times[\vort\times(\vnvec-\vsvec)].
\label{eq:fdef}
\ee
For getting a qualitative feeling for the relative sizes of the
phenomenological quantities $\beta$ and $\betap$, we use a 
different parametrization
for the strength of the mutual friction force, based on the idea of
vortex drag. The equation for the superfluid velocity including
mutual friction is
\be
{\partial\vsvec\over\partial t}+\vsvec\cdot\nabla\vsvec
=-\nabla(\mu+\phi)+{{\bf f}\over\rhos},
\ee
where $\mu$ is the chemical potential and $\phi$ the gravitational
potential; taking the curl of this equation gives
\be
{\partial\vort\over\partial t}=\curl(\vsvec\times\vort+{\bf f}/\rhos).
\ee
If ${\bf f}=0$, then the superfluid vortex lines comove with the
superfluid, but, in general, the vortex lines have a different
velocity, $\vlvec\neq\vsvec$, and
\be
{\partial\vort\over\partial t}=\curl(\vlvec\times\vort);
\ee
from the form for $\bf f$ given in equation (\ref{eq:fdef}), we can read off
\be
\vlvec=\vsvec+\betap(\vnvec-\vsvec)+\beta\vorthat\times(\vnvec-
\vsvec)
=\vnvec+(\betap-1)(\vnvec-\vsvec)+\beta\vorthat\times(\vnvec-
\vsvec).
\ee
Only the components of $\vnvec-\vsvec$ perpendicular to $\vorthat$
contribute to $\vlvec$, as can be seen from the original expression for $\bf f$.
Clearly, vortex lines comove with the superfluid if $\vert\betap\vert$ and $\beta$
are both small, and comove with the normal fluid if $\vert\betap-1\vert$ and $\beta$
are small. (If superfluid rotates faster than normal fluid, vortices move slowly
outward relative to normal fluid for $\beta$ small.)
The motion of a vortex is found by balancing the Magnus force
due to superfluid streaming past the line and any other forces it experiences;
for our purposes, the latter are drag forces perpendicular to the line, so
the equation of motion is 
\be
\rhos\kappa\vorthat\times(\vlvec-\vsvec)+\fdrag=0.
\ee
If the drag force per length on a vortex is
\be
\fdrag=-\eta(\vlvec-\vnvec)-\etap\vorthat\times(\vlvec-\vnvec),
\ee
then the vortex line velocity is
\be
\vlvec=\vsvec+{[\eta^2-\etap(\rhos\kappa-\etap)](\vnvec-\vsvec)\over
(\rhos\kappa-\etap)^2+\eta^2}
+{\eta\kappa\rhos\vorthat\times(\vnvec-\vsvec)\over
(\rhos\kappa-\etap)^2+\eta^2},
\ee
from which we infer the relations
\be
\beta={\eta\rhos\kappa\over(\rhos\kappa-\etap)^2+\eta^2}.
\label{eq:betadef} 
\ee
\be
\betap=1-{\rhos\kappa(\rhos\kappa-\etap)\over
(\rhos\kappa-\etap)^2+\eta^2}.
\label{eq:betapdef}
\ee
These results relate the drag coefficients $\eta$ and $\etap$
with the parameters $\beta$ and $\betap$ appearing in the mutual
friction force. 

In microscopic models for mutual friction developed
so far, the
coefficient $\etap$, which determines the magnitude of the
drag force perpendicular to the motion of a vortex line through
the normal fluid, is negligible. If $\etap=0$, then equations
(\ref{eq:betadef}) and (\ref{eq:betapdef}) simplify to
\be
\beta={\eta\rhos\kappa\over(\rhos\kappa)^2+\eta^2}.
\label{eq:betadefzero}
\ee
\be
\betap=1-{(\rhos\kappa)^2\over
(\rhos\kappa)^2+\eta^2}
\label{eq:betapdefzero}.
\ee
From these relationships, we find that when $\eta\gg\rhos\kappa$
vortex lines are dragged effectively, and tend to follow the
normal fluid closely; in that limit,
$\beta\approx
\rhos\kappa/\eta$ and $1-\betap\approx(\rhos\kappa/\eta)^2
\approx\beta^2$. When $\eta\ll\rhos\kappa$ the drag is weak, 
vortex lines tend to follow the superfluid, and
$\beta\approx\eta/\rhos\kappa$ and $\betap\approx
(\eta/\rhos\kappa)^2\approx\beta^2$. As we shall see below,
this means that the dissipative torque arising from mutual
friction is much larger than the nondissipative torque when
the drag is either very strong or very weak; the two torques
are only comparable when $\eta/\rhos\kappa\sim 1$. When
$\etap\neq 0$, the situation becomes more complicated, as
equations (\ref{eq:betadef}) and (\ref{eq:betapdef})
involve two nondimensional parameters, $\eta/\rhos/\kappa$
and $\etap/\rhos\kappa$. If we assume that $\etap\ll\eta$,
then in the strong damping limit, $\beta\approx\rhos\kappa/
\eta$ as before, and $1-\betap\approx(\rhos\kappa/\eta)^2$
if $\etap\ll\rhos\kappa$, but $1-\betap\approx -\etap\rhos
\kappa/\eta\approx -(\etap/\eta)\beta$. In the weakly 
coupled domain, $\beta\approx\eta/\rhos\kappa$ as before,
but $\betap\approx\beta^2$ only when $\etap/\eta\ll\eta/\rhos
\kappa$, and instead $\betap\approx -\etap/\rhos\kappa
\approx -(\etap/\eta)\beta$ when $\etap/\eta\gg\rhos\kappa$.
As we shall see, these results will make the existence of
long term oscillatory modes problematic when $\etap\ll\eta$
provided that at least part of the crust is coupled strongly
to the crustal superfluid.
For $\etap\simgreat\eta$, the situation turns out to be
more favorable 
for the survival of oscillatory modes. In that case,
$\beta\approx\eta\rhos\kappa/
{\etap}^2$ and $1-\betap\approx -\rhos\kappa/\etap\approx
-(\etap/\eta)\beta$ in the limit of strong coupling,
and $\beta\approx\eta/\rhos\kappa$ and 
$\betap\approx -\etap/\rhos\kappa\approx -(\etap/\eta)\beta$
in the limit of weak coupling.

The torque that results from mutual friction is
\be
\Nvec=\int{d^3\rvec\,\rvec\times\fvec(\rvec)}\equiv\nvec+\nvecp,
\ee
where, from equation (\ref{eq:fdef}),
\be
\nvec=-\int{d^3\rvec\,\beta\rhos\rvec\times\{
\vorthat\times[\vort\times(\vnvec-\vsvec)]\}}
\label{eq:nvecdef}
\ee
\be
\nvecp=-\int{d^3\rvec\,\betap\rhos\rvec\times
[\vort\times(\vnvec-\vsvec)]}.
\label{eq:nvecpdef}
\ee
We restrict ourselves to a uniformly rotating normal fluid,
but the analogous restriction to uniformly rotating superfluid
is dynamically inconsistent unless $\beta$ and $\betap$ are
independent of position. Consequently, we imagine that the
star can be divided into ``shells'' in which $\beta$ and
$\betap$ are independent of position, and the superfluid
rotates uniformly. In these shells, 
$\vsvec=\Omsvec\times\rvec$ and
$\vnvec=\Omnvec\times\rvec$, with $\Omnvec$ and $\Omsvec$
independent of $\rvec$; this also implies that $\vort
=2\Omsvec$. In succeeding sections, we consider stars with
one and two superfluid shells. These examples suffice to
illustrate the complex behavior that may arise in a real neutron
star, where $\beta$ and $\betap$ vary continuously.

For uniform rotation, 
equations (\ref{eq:nvecdef}) and (\ref{eq:nvecpdef}) become
\be
\nvec=\Omsvec\times\Tbeta\cdot({\hat\Omsvec}\times\Omnvec)
+\Omega_s(\Omsvec-\Omnvec)\cdot[\Tbeta-\delvec\Tr(\Tbeta)]
\label{eq:nveceval}
\ee
\be
\nvecp=-(\Omsvec-\Omnvec)\times\Tbetap\cdot\Omsvec,
\label{eq:nvecpeval}
\ee
where 
\be
\Tbeta\equiv 2\beta\int{d^3\rvec\rhos\rvec\rvec},
\label{eq:tbetadef}
\ee
\be
\Tbetap\equiv 2\betap\int{d^3\rvec\rhos\rvec\rvec},
\label{eq:tbetapdef}
\ee
$\delvec$ is the unit tensor, and $\Tr(\Tbeta)$ is the
trace of $\Tbeta$.
It is easy to show that $(\Omsvec-\Omnvec)\cdot\nvecp=0$
and $(\Omsvec-\Omnvec)\cdot\nvec<0$, so that 
$(\Omsvec-\Omnvec)\cdot\Nvec<0$, that is, mutual friction
torques are ultimately dissipative.

The tensors $\Tbeta$ and $\Tbetap$ can be rather complicated
in general. Even in uniformly rotating superfluid shells,
the superfluid density $\rhos(\rvec)$ may be slightly anisotropic,
principally as a result of rotational flattening perpendicular
to $\Omsvec$, which is time varying and not aligned 
with any of the principal axes of the crust in general. However, we
shall neglect these complications, since 
the magnitudes of the anisotropies in $\Tbeta$ and $\Tbetap$
are expected to be small for slowly rotating neutron stars,
which we focus on here. Accordingly, we approximate
\be
\Tbeta=I_s\beta_{\rm eff}\delta
\ee
\be
\Tbetap=I_s\betap_{\rm eff}\delta,
\ee
where $I_s$ is the moment of inertia of the superfluid, and
$\beta_{\rm eff}$ and $\betap_{\rm eff}$ are suitably averaged
$\beta$ and $\betap$; henceforth, we drop the subscript ``eff.''
With these expressions for $\Tbeta$ and $\Tbetap$ the mutual
friction torques simplify to
\be
\nvec=-I_s\beta\Omega_s(\Omsvec-\Omnvec)\cdot(\delvec+\hat
\Omsvec\hat\Omsvec)
\label{eq:nvecfinal}
\ee
\be
\nvecp=I_s\betap(\Omnvec\times\Omsvec).
\label{eq:nvecpfinal}
\ee
We shall devote much of the remainder of this paper to examining
the consequences of torques of this form. Here, we neglect other
torques which could be important, such as gravitational 
torques (both Newtonian and post-Newtonian) or fluid torques arising 
from boundary conditions, and ignore the
various complications in $\Tbeta$ and $\Tbetap$ alluded to above.
Some of these issues will be discussed in a subsequent publication.

\section{TWO COMPONENT STAR}\label{sec:twocomp}

Implicit in the review of Shaham (1977) presented in Section
\ref{sec:sh77} was a treatment of the two component system
consisting of the rigid crust and pinned crustal superfluid.
This was the $\gamma=0$ limit in which the crust and core
decouple entirely. In that case, we found that the crust
precesses at a frequency $\sigma$ (see eq. [\ref{eq:sigdef}])
under the additional assumption of axisymmetry; for $\gamma
\equiv 0$, this mode is undamped. The three component model
discussed in Section \ref{sec:sh77} also reduces to a
two component system when $\gamma\to\infty$, in which case
the core and crust are coupled perfectly, and must corotate.
Although this limit is not realistic (see discussion of the
maximum possible $\gamma$ in the penultimate paragraph
of Section \ref{sec:sh77}), it also leads to undamped 
precession at a frequency $\sigma I_c/(I_1+I_c)$. Here,
we examine how imperfect pinning alters these results, and
introduces both damping and new modes.

\subsection{Free Precession Reexamined}\label{sec:freeprec}

The coupled equations for the angular momenta of the crust
and crustal superfluid are
\be
{d(\Icr\cdot\Omcr)\over dt}=-\nvec-\nvecp
=I_s\beta\Omega_s(\Omsvec-\Omcr)\cdot(\delvec+\hat\Omsvec
\hat\Omsvec)+I_s\betap(\Omsvec\times\Omcr)
\label{eq:lcr}
\ee
\be
I_s{d\Omsvec\over dt}=\nvec+\nvecp
=-I_s\beta\Omega_s(\Omsvec-\Omcr)\cdot(\delvec+\hat\Omsvec
\hat\Omsvec)+I_s\betap(\Omcr\times\Omsvec),
\label{eq:ls}
\ee
where we have substituted $\Omcr$ for $\Omnvec$ in equations
(\ref{eq:nvecfinal}) and (\ref{eq:nvecpfinal}), and assumed
that the angular momentum of the superfluid is $I_s\Omsvec$,
which is tantamount to assuming that the moment of inertia
tensor of the superfluid is of the form
\be
{\bf I_s}=(I_s-\Delta I_s)\delvec+
\Delta I_s\biggl({3\hat\Omsvec\hat\Omsvec-\delvec\over 2}
\biggr).
\ee
Notice that if $\betap=1$ and $\beta=0$, these equations
reduce to
\be
{d(\Icr\cdot\Omcr)\over dt}+I_s(\Omcr\times\Omsvec)=0,
\ee
which is equivalent to equation (\ref{eq:angmom1}) with 
the contribution from the core component omitted,
and
\be
{d\Omsvec\over dt}=\Omcr\times\Omsvec,
\ee
which implies that $\Omsvec$ is fixed in the reference frame
that rotates with the superfluid. This is the limit of
perfect pinning, and results in undamped precession at 
the frequency $\sigma$.

When $\betap\neq 1$ and $\beta\neq 0$, there are additional
modes. Let us work in the frame rotating with the crust,
in which case (recall that $d^\star{\bf F}/dt$ is the time
derivative of $\bf F$ in this frame)
\be
\Icr\cdot{d^\star\Omcr\over dt}+\Omcr\times(\Icr\cdot\Omcr)
=I_s\beta\Omega_s(\Omsvec-\Omcr)\cdot(\delvec+\hat\Omsvec
\hat\Omsvec)+I_s\betap(\Omsvec\times\Omcr)
\label{eq:lcr1}
\ee
\be
{d^\star\Omsvec\over dt}+(1-\betap)(\Omcr\times\Omsvec)
=-\beta\Omega_s(\Omsvec-\Omcr)\cdot(\delvec+\hat\Omsvec
\hat\Omsvec).
\label{eq:oms1}
\ee
If we project equations (\ref{eq:lcr1}) and (\ref{eq:oms1})
along the principal axes of the crust, and linearize around
the fixed point at which $\Omcr=\Omsvec=\Omvec$ and $\Omvec
\parallel\hat{\bf e_3}$ we find
\be
\dot\Omcrone+\biggl[\biggl({I_3-I_2\over I_1}\biggr)
+{I_s\betap\over I_1}\biggr]\Omcrtwo
-{I_s\betap\over I_1}\Omstwo
=-{I_s\beta\over I_1}(\Omcrone-\Omsone)
\label{eq:omcrone}
\ee
\be
\dot\Omcrtwo-\biggl[\biggl({I_3-I_1\over I_2}\biggr)
+{I_s\betap\over I_2}\biggr]\Omcrone
+{I_s\betap\over I_2}\Omsone
=-{I_s\beta\over I_2}(\Omcrtwo-\Omstwo)
\label{eq:omcrtwo}
\ee
\be
\dot\Omcrthree=-2{I_s\beta\over I_3}(\Omcrthree-\Omsthree)
\ee
\be
\dot\Omsone-(1-\betap)(\Omstwo-\Omcrtwo)=
-\beta(\Omsone-\Omcrone)
\label{eq:omsone}
\ee
\be
\dot\Omstwo+(1-\betap)(\Omsone-\Omcrone)=
-\beta(\Omstwo-\Omcrtwo)
\label{eq:omstwo}
\ee
\be
\dot\Omsthree=-2\beta(\Omsthree-\Omcrthree).
\ee
As before, the evolution of the perturbations along the 3-axis
decouple from those along the other axes, and decay exponentially;
the rate of decay is $2\beta(1+I_s/I_3)$. The remaining equations
imply a fourth order characteristic equation if we search for
modes $\propto\exp(p\phi)$.

\subsubsection{Axisymmetric Crust}\label{sec:axi}

When the crust is axisymmetric, $I_1=I_2$, and equations (\ref{eq:omcrone}),
and (\ref{eq:omcrtwo}) simplify to
\be
\dot\Omcrone+\biggl[\biggl({I_3-I_1\over I_1}\biggr)
+{I_s\betap\over I_1}\biggr]\Omcrtwo
-{I_s\betap\over I_1}\Omstwo
=-{I_s\beta\over I_1}(\Omcrone-\Omsone)
\ee
\be
\dot\Omcrtwo-\biggl[\biggl({I_3-I_1\over I_1}\biggr)
+{I_s\betap\over I_1}\biggr]\Omcrone
+{I_s\betap\over I_1}\Omsone
=-{I_s\beta\over I_1}(\Omcrtwo-\Omstwo);
\ee
these couple to 
equations (\ref{eq:omsone}) and (\ref{eq:omstwo}), which are unchanged.

As we found in Section \ref{sec:sh77}, the fourth order characteristic
equation may be reduced to second order in this case. Define
\be
\sigmap=\biggl({I_3-I_1\over I_1}\biggr)+{I_s\betap\over I_1}
\equiv\epsilon+{I_s\betap\over I_1}
\label{eq:sigmapdef}
\ee
and let
\be
\Omstil=\Omsone+i\Omstwo;
\ee
then we get the two coupled equations
\be
\dot\Omcrtil-i\sigmap\Omcrtil+i{I_s\betap\over I_1}\Omstil
=-{I_s\beta\over I_1}(\Omcrtil-\Omstil)
\ee
\be
\dot\Omstil+i(1-\betap)(\Omstil-\Omcrtil)=
-\beta(\Omstil-\Omcrtil).
\ee
It turns out to be convenient to use 
\be
\Delomtil\equiv \Omstil-\Omcrtil
\ee
instead of $\Omstil$; doing so yields the coupled equations
\be
\dot\Omcrtil-i\epsilon\Omcrtil+{I_s\over I_1}(i\betap-\beta)
\Delomtil=0
\label{eq:axiomcr}
\ee
and
\be
\dot\Delomtil+\biggl\{i\biggl[1-\betap\biggl(1+{I_s\over I_1}
\biggr)\biggr]+\beta\biggl(1+{I_s\over I_1}\biggr)\biggr\}
\Delomtil
+i\epsilon\Omcrtil=0.
\label{eq:axidelom}
\ee
Assuming that $(\Omcrtil,\Delomtil)\propto\exp(p\phi)$ we
find the relation
\be
\Delomtil=-{p\Omcrtil\over p+i(1-\betap)+\beta},
\label{eq:aximode}
\ee
and the characteristic equation
\be
p^2+p\biggl[i(1-\betap-\sigmap)+\beta\biggl(1+{I_s\over I_1}\biggr)
\biggr]+\epsilon(1-\betap-i\beta)=0.
\label{eq:axichar}
\ee
Equation (\ref{eq:aximode}) is useful
for finding the eigenvectors once equation (\ref{eq:axichar})
is solved; these are needed to determine the response of the two
spin components to external torques.

Although we can solve equation (\ref{eq:axichar}) exactly, it is
instructive to consider the two limiting cases of weak and strong
vortex drag separately. When vortex drag is weak, $\beta$ and
$\betap$ are small in magnitude, so we rewrite equation
(\ref{eq:axichar}) as
\be
(p-i\epsilon)(p+i)-(\betap+i\beta)\biggl[ip\biggl(1+{I_s\over I_1}
\biggr)+\epsilon\biggr]=0.
\ee
The solutions to this
equation to first order in the
small quantities $\beta$ and $\betap$ are	
\be
p_d=-i+{(i\betap-\beta)(1+\epsilon+I_s/I_1)\over 1+\epsilon}
\label{eq:pdbwk}
\ee
and (the Shaham mode)
\be
p_p=i\epsilon+{\epsilon(i\betap-\beta) I_s\over I_1(1+\epsilon)}.
\label{eq:axiprecweak}
\ee
Both of these solutions damp slowly, at rates proportional to
$\beta>0$. The second mode reduces to the conventional Euler
precession when $\beta=\betap=0$. The first mode arises because the
superfluid angular velocity would remain fixed in the inertial frame
if $\beta$ and $\betap$ were zero, but wanders slowly when the 
coupling is small but nonzero. (We discuss this point more fully
in the context of three component models; see Section \ref{sec:crcscor} below,
discussion following eq. [\ref{eq:pdpbstrzwk}].)
For strong vortex drag,
$1-\betap$ and $\beta$ are small, and we rewrite equation 
(\ref{eq:axichar}) in the form
\be
p(p-i\sigmap)+(1-\betap)(ip+\epsilon)
+\beta\biggl[p\biggl(1+{I_s\over I_1}\biggr)-i\epsilon\biggr]
=0.
\ee
The solutions to this equation to first order in the small
quantities $\beta$ and $1-\betap$ are
\be
p_d=-{\epsilon[\beta+i(1-\betap)]
\over\sigmap}
\label{eq:pdbstr}
\ee
and
\be
p_p=i\sigmap-{I_s[i(1-\betap)+\beta(1+\sigmap)]\over I_1\sigmap}.
\label{eq:axiprecstrong}
\ee
The first of these solutions represents a slowly damped mode with
an oscillatory part that is negligible when $\etap\ll\eta$,
implying $1-\betap\ll\beta$ in this limit. (Recall discussion
following eqs. [\ref{eq:betadef}] and [\ref{eq:betapdef}] in
Section \ref{sec:setup}.)
The second mode corresponds to precession at $\sigmap$
with slow damping: For $I_s/I_1\gg\epsilon$, equation (\ref{eq:sigmapdef})
implies that $\sigmap\approx I_s/I_1$ for $1-\betap\ll 1$, so the
damping rate is approximately $\beta(1+I_s/I_1)$; in the unlikely
event that $\epsilon\gg I_s/I_1$, then $\sigmap\approx\epsilon$
and the damping rate is approximately $(I_s/I_1\epsilon)\beta
(1+\epsilon)$.

\subsubsection{Non-Axisymmetric Crust}\label{sec:nonaxi}

Since the neutron star crust may not be axisymmetric, it is worth
checking that there are no surprises when $I_1\neq I_2$. Define
\be
\sigmap_1=\biggl({I_3-I_2\over I_1}\biggr)+{I_s\betap\over I_1}
\equiv\epsilon_1+{I_s\betap\over I_1}
\ee
\be
\sigmap_2=\biggl({I_3-I_1\over I_2}\biggr)+{I_s\betap\over I_2}
\equiv\epsilon_2+{I_s\betap\over I_2};
\ee
then equations (\ref{eq:omcrone}) and (\ref{eq:omcrtwo}) become
\be
\dot\Omcrone+\sigmap_1\Omcrtwo-{I_s\betap\over I_1}\Omstwo
=-{I_s\beta\over I_1}(\Omcrone-\Omsone)
\ee
\be
\dot\Omcrtwo-\sigmap_2\Omcrone+{I_s\betap\over I_2}\Omsone
=-{I_s\beta\over I_2}(\Omcrtwo-\Omstwo),
\ee
which must be solved along with equations
(\ref{eq:omsone}) and (\ref{eq:omstwo}).
When we look for solutions $\propto\exp(p\phi)$ we find the
fourth order characteristic equation
\begin{eqnarray}
0 & = &
p^4+p^3\beta\biggl(2+{I_s\over I_1}+{I_s\over I_2}\biggr)
\nonumber\\ & &
+p^2\biggl\{[\beta^2+(1-\betap)^2]\biggl(1+{I_s\over I_1}\biggr)
\biggl(1+{I_s\over I_2}\biggr)-(1-\betap)^2{I_s^2\over I_1I_2}
-(1-\betap)\biggl({I_s\over I_1}+{I_s\over I_2}\biggr)
+\sigmap_1\sigmap_2\biggr\}
\nonumber\\ & &
+p\beta\biggl(2\epsilon_1\epsilon_2
+{I_s\epsilon_2\over I_1}+{I_s\epsilon_1\over I_2}\biggr)
+\epsilon_1\epsilon_2[\beta^2+(1-\betap)^2].
\label{eq:gruesome}
\end{eqnarray}
It is not possible to factorize the characteristic equation into
the product of two second order equations because there is no
guarantee that all of the roots are simply complex conjugate
pairs.

In the limit of weak coupling, expanding equation (\ref{eq:gruesome})
up to first order in $\beta$ and $\betap$ yields
\begin{eqnarray}
0& = &(p^2+1)(p^2+\epsone\epstwo)
+\beta\biggl[p^3\biggl(2+{I_s\over I_1}+{I_s\over I_2}\biggr)
+p\biggl(2\epsone\epstwo+{I_s\epstwo\over I_1}+{I_s\epsone\over I_2}
\biggr)\biggr]
\nonumber\\ & &
-\betap\biggl\{p^2\biggl[2+{I_s(1-\epstwo)\over I_1}+
{I_s(1-\epsone)\over I_2}\biggr]+2\epsone\epstwo\biggr\}.
\end{eqnarray}
The approximate solutions of this form for the characteristic equation
are
\be
p_d=-i
+{[2(1-\epsone\epstwo)+(I_s/I_1)(1-\epstwo)+(I_s/I_2)(1-\epsone)]
(i\betap-\beta)\over 2(1-\epsone\epstwo)}
\ee
and
\begin{eqnarray}
p_p & = & i\sqrt{\epsone\epstwo}
+{i\betap\sqrt{\epsone\epstwo}
[(I_s/I_1)(1-\epstwo)+(I_s/I_2)(1-\epsone)]\over 2(1-\epsone\epstwo)}
\nonumber\\ & & 
-{\beta[(I_s/I_1)\epstwo(1-\epsone)
+(I_s/I_2)\epsone(1-\epstwo)]\over 2(1-\epsone\epstwo)},
\end{eqnarray}
and their complex conjugates.
To get limiting results in the strongly coupled domain, we rewrite
equation (\ref{eq:gruesome}) in the slightly modified form
\begin{eqnarray}
0 & = & p^2(p^2+\sigmap_1\sigmap_2)+
p^3\beta\biggl(2+{I_s\over I_1}+{I_s\over I_2}\biggr)
\nonumber\\ & &
+p^2\biggl\{[\beta^2+(1-\betap)^2]\biggl(1+{I_s\over I_1}\biggr)
\biggl(1+{I_s\over I_2}\biggr)-(1-\betap)^2{I_s^2\over I_1I_2}
-(1-\betap)\biggl({I_s\over I_1}+{I_s\over I_2}\biggr)
\biggr\}
\nonumber\\ & &
+p\beta\biggl(2\epsilon_1\epsilon_2
+{I_s\epsilon_2\over I_1}+{I_s\epsilon_1\over I_2}\biggr)
+\epsilon_1\epsilon_2[\beta^2+(1-\betap)^2].
\label{eq:gruesome1}
\end{eqnarray}
From equation (\ref{eq:gruesome1}), it is evident that the 
modes are near $p^2=0$ and $p^2=-\sigmap_1\sigmap_2$. To get the
first order approximation to the modes with $p^2=0$ to zeroth
order in $\beta$ and $1-\betap$, we identify all terms in 
equation (\ref{eq:gruesome1}) that are potentially second order
in small quantities; this leads to the quadratic equation
\be
0=p^2\sigmap_1\sigmap_2+
p\beta\biggl(2\epsone\epstwo+{I_s\epstwo\over I_1}+{I_s\epsone
\over I_2}\biggr)+\epsone\epstwo[\beta^2+(1-\betap)^2],
\ee
which has the pair of roots
\be
p_d^\pm=-{\beta(\epsone\sigmap_2+\epstwo\sigmap_1)\over
2\sigmap_1\sigmap_2}
\pm
\biggl[{\beta^2(\epsone\sigmap_2-\epstwo\sigmap_1)^2\over 
(2\sigmap_1\sigmap_2)^2}
-{\epsone\epstwo(1-\betap)^2\over\sigmap_1\sigmap_2}
\biggr]^{1/2}.
\label{eq:pdnonaxi}
\ee
In the axisymmetric limit, these two roots reduce to the
damped mode found in Section \ref{sec:axi} and its complex
conjugate, but although $p_d^\pm$ both imply damping in general,
they could be purely real and different in magnitude,
especially since we expect $1-\betap$ to be much smaller
than $\beta$ in the strongly coupled regime. (This is 
why we could not factor eq. [\ref{eq:gruesome}] into two
quadratic equations.) It is also straightforward to expand
equation (\ref{eq:gruesome1}) around the approximate root
$p\approx i\sqrt{\sigmap_1\sigmap_2}$ to find
\be
p_p=i\sqrt{\sigmap_1\sigmap_2}
-{i(1-\betap)\over 2\sqrt{\sigmap_1\sigmap_2}}
\biggl({I_s\over I_1}+{I_s\over I_2}\biggr)
-{\beta\over 2\sqrt{\sigmap_1\sigmap_2}}
\biggl[{I_s\over I_1}\sigmap_2(1+\sigmap_1)
+{I_s\over I_2}\sigmap_1(1+\sigmap_2)\biggr],
\ee
to first order in the small quantities $\beta$ and
$1-\betap$.

From this brief foray into the modes of a triaxial star, we conclude
that deviations from axisymmetry do not alter the behavior
of the precession qualitatively
in either limit. The character of the damped modes may be
different in the strong coupling limit for nonaxisymmetric
stars, but if so, they become purely damped, with
no oscillation at all. Consequently, from here on we 
specialize to axisymmetric crusts, since we do not expect to
miss any important oscillatory modes.

\subsection{Response to External Torques}\label{sec:torq2}

As we have seen, the modes of free precession for this two component
model are rapidly oscillating and/or damped. Here, we consider the
response of the system to external torques. Our analysis will reveal
that the limit of perfect coupling between the normal fluid and
superfluid must be taken with care when external torques act.

If we suppose that the crust is subject to an arbitrary time-dependent
torque, $\Nvec_{cr}(\phi)$, then equations (\ref{eq:axiomcr}) and
(\ref{eq:axidelom}) are changed to
\be
\dot\Omcrtil-i\epsilon\Omcrtil+{I_s\over I_1}(i\betap-\beta)
\Delomtil=\Nveccr(\phi)
\ee
and
\be
\dot\Delomtil+\biggl\{i\biggl[1-\betap\biggl(1+{I_s\over I_1}
\biggr)\biggr]+\beta\biggl(1+{I_s\over I_1}\biggr)\biggr\}
\Delomtil
+i\epsilon\Omcrtil=-\Nveccr(\phi),
\ee
where $\Nveccr\equiv I_1^{-1}(\Nvec_{cr,1}+i\Nvec_{cr,2})$.
Apart from decaying transients, the solution to these equations is
\be
\Omcrtil=\sum_{\alpha=p,d}A_\alpha
\int_{-\infty}^\phi{d\phip\Nveccr(\phip)
\exp[p_\alpha(\phi-\phip)]}
\label{eq:responsomcr}
\ee
\be
\Delomtil=-\sum_{\alpha=p,d}{p_\alpha A_\alpha\over
p_\alpha+\Bvar}
\int_{-\infty}^\phi{d\phip\Nveccr(\phip)
\exp[p_\alpha(\phi-\phip)]},
\label{eq:responsedelom}
\ee
where the coefficients are
\be
A_p={[p_p+\Bvar]\over (p_p-p_d)}\qquad
A_d={[p_d+\Bvar]\over (p_d-p_p)}.
\ee
(Equation [\ref{eq:aximode}] is useful in obtaining the
$A_\alpha$.)

In the limit of perfect coupling between the crust and
superfluid, we take $\Bvar\to 0$ with
$p_d/[\Bvar]=-\epsilon/\sigmap$ constant; taking this
limit of equations (\ref{eq:responsomcr}) and
(\ref{eq:responsedelom}) na\"ively yields
\be
\Omcrtil=\int_{-\infty}^\phi{d\phip\Nveccr\exp[p_p(\phi-\phip)]}
\label{eq:perfect}
\ee
with $\Delomtil\to-\Omcrtil$.

Actually, the perfect coupling limit is a bit more subtle than
the manipulations leading to equation (\ref{eq:perfect}). To see
why, consider the response to a time-independent torque on
the crust. Then equation (\ref{eq:responsomcr}) may be integrated
readily and we find 
\be
\Omcrtil={i\Nveccr\over\epsilon};
\label{eq:constorq}
\ee
integrating equation (\ref{eq:responsedelom}) yields $\Delomtil=0$.
These results ought to hold for {\it any} $\Bvar$. However,
equation (\ref{eq:perfect}), which purports to describe the
limit of perfect coupling, $\Bvar\equiv 0$,
yields
\be
\Omcrtil=-{\Nveccr\over p_p}\to {i\Nveccr\over\sigmap}
\label{eq:constorqperf}
\ee
(assuming a small, negative real part to $p_p$). Which of these
results is correct?

To resolve the conundrum, consider a torque that turns on (or can
be regarded as constant since) some time in the past, $\phi_0$.
Then equations (\ref{eq:responsomcr})
and (\ref{eq:responsedelom}) yield the response
\be
\Omcrtil=\biggl\{{i\over\epsilon}+
\biggl(1+{\Bvar\over p_p}\biggr){\exp[p_p(\phi-\phi_0)]\over
p_p-p_d}
-\biggl(1+{\Bvar\over p_d}\biggr){\exp[p_d(\phi-\phi_0)]\over
p_p-p_d}\biggr\}\Nveccr
\ee
\be
\Delomtil=\biggl\{{-\exp[p_p(\phi-\phi_0)]+\exp[p_d(\phi-\phi_0)]\over
p_p-p_d}\biggr\}\Nveccr.
\ee
We suppose that the damping constant associated with the precessing
mode, $p_p$, is large enough that $\exp[p_p(\phi-\phi_0)]\to 0$;
then what we find depends on $p_d(\phi-\phi_0)$. (Recall that the
real parts of $p_p$ and $p_d$ are negative.) If 
$\vert {\rm Re}[p_d(\phi-\phi_0)]\vert
\gg 1$, so that any transient response has plenty of time to damp
away between $\phi_0$ and $\phi$, then we recover equation 
(\ref{eq:constorq}), and also find that $\Delomtil\to 0$. On the
other hand, if 
$\vert{\rm Re}[p_d(\phi-\phi_0)]\vert\ll 1$, then we recover equation
(\ref{eq:constorqperf}) and $\Delomtil\to -\Omcrtil$ in the
limit of perfect coupling, $\Bvar\to 0$, to zeroth order in
$p_d(\phi-\phi_0)$. To first order, we also find a term that
grows linearly with $p_d(\phi-\phi_0)$; over a sufficiently long
timespan, this growth would change the tilt from equation
(\ref{eq:constorq}) to equation (\ref{eq:constorqperf}).

The resolution of the apparent paradox is that there is none:
equations (\ref{eq:responsomcr}) and (\ref{eq:responsedelom})
are always the correct ones to use. What one gets in the limit
of strong vortex drag depends on how the timescale on which
the external torque changes compares with the timescales
inherent in the coupling of superfluid to the normal crust.
Equations (\ref{eq:constorq}) and (\ref{eq:constorqperf}) both have
domains of validity; equation (\ref{eq:constorqperf}) is a lower
bound to the steady-state tilt of the rotational angular velocity away from
the 3-axis in the strong coupling limit. As long as the
damping timescale associated with $p_d$ is short compared to
any timescale associated with changes in the external torque,
however, equation (\ref{eq:constorq}) gives the right
response. In practical terms, pulsar spindown provides a nearly
constant torque on the crust which can give rise to $\Nveccr$.
Thus, if $p_d$ implies decay on timescales smaller than the
pulsar spindown time, then equation (\ref{eq:constorq}) describes
the response of the crust, even in the strong pinning limit.

We emphasize that this result could not be found from a consideration
of normal modes alone; arriving at it requires examining the response
of the star to a torque. There is therefore a subtle aspect to 
the limiting case considered by Shaham (1977): While the modal
frequencies he derived are correct, and his conclusions about 
free precession warranted, blithely using equation (\ref{eq:perfect}),
which would follow from the assumption of perfect pinning,
instead of equations (\ref{eq:responsomcr}) and (\ref{eq:responsedelom})
is wrong {\it even} in the limit of strong pinning. 

The response of the star to torques along the 3-axis is
found by solving the equations
\be
\dot\Omcrthree-{2I_s\beta\over I_3}\Delomthree=\Nveccrthree
(\phi)
\ee
and
\be
\dot\Delomthree+2\beta\biggl(1+{I_s\over I_3}\biggr)
\Delomthree
=-\Nveccrthree(\phi),
\ee
where $\Nveccrthree=\Nvec_{cr,3}/I_3$;
the result is
\be
\Delomthree=-\int_{-\infty}^\phi{d\phip\Nveccrthree(\phip)
\exp\biggl[-2\beta\biggl(1+{I_s\over I_3}\biggr)(\phi-\phip)
\biggr]}
\label{eq:impcrncr}
\ee
and
\be
\Omcrthree={I_3\over I_3+I_s}\int_{-\infty}^\phi
{d\phip\Nveccrthree(\phip)}
+{I_s\over I_s+I_3}\int_{-\infty}^\phi
{d\phip\Nveccrthree(\phip)
\exp\biggl[-2\beta\biggl(1+{I_s\over I_3}\biggr)(\phi-\phip)
\biggr]}.
\label{eq:impsfncr}
\ee
In particular, a time-independent torque results in a steady
angular velocity difference
\be
\Delomthree=-{\Nveccrthree\over 2\beta(1+I_s/I_3)},
\ee
while the crustal angular velocity changes linearly:
\be
\dot\Omcrthree={\Nveccrthree I_3\over I_s+I_3}.
\ee
The response to an impulsive torque $\Nveccrthree=
\Nvec_{0,3}\delta(\phi-\phi_0)$ is
\be
\Omcrthree={\Nvec_{0,3}\{I_3+I_s\exp[-2\beta(1+I_s/I_3)(\phi
-\phi_0)]\}\over I_s+I_3}
\ee
\be
\Delomthree=-\Nvec_{0,3}\exp[-2\beta(1+I_s/I_3)(\phi
-\phi_0)].
\ee

Spindown torques, which only change on very long timescales
for all observed pulsars, provide a physical realization of
a ``time-independent'' torque. Let us take the vacuum
magnetic dipole torque, which is proportional to 
$-\muvec\times(\omvec\times\muvec)$ (e.g. Davis and Goldstein
1970, Goldreich 1970, Michel 1991); assuming a magnetic
dipole moment $\muvec$ that is fixed in the frame of the
crust, with components
\be
\muvec=\cos\alpha{\bf\hat e_3}+\sin\alpha{\bf\hat e_1},
\ee
the torque may be written (to a good first
approximation) as
\be
\Nvec_{d}=N_d\sin\alpha({\bf\hat e_1}\cos\alpha
-{\bf\hat e_3}\sin\alpha)\equiv
N_{sd}({\bf\hat e_1}\cot\alpha-{\bf\hat e_3}),
\ee
implying $\Nveccr=N_{sd}\cot\alpha/I_1$ and 
$\Nveccrthree=-N_{sd}/I_3$. 
The steady state response to these torques is
\be
\Omcrtil={iN_{sd}\cot\alpha\over\epsilon I_1}
\qquad{\rm and}\qquad
\Delomtil=0
\ee
and
\be
\Delomthree={N_{sd}\over 2\beta(I_s+I_3)},
\ee
with $\dot\Omcrthree=-N_{sd}/I_3$. Two aspects of these
results are especially noteworthy. First, as is
well-known, the superfluid rotates faster than the
crust as a consequence of the spindown torque.
Second, but not so widely appreciated, the steady
state ``tilt'' in the angular velocity of the crust
is surprisingly large, as it is proportional to 
$\epsilon^{-1}$. (When $\epsilon=0$, eq.
[\ref{eq:axichar}] implies that $p_d=0$, and
$\Omcrtil$ grows linearly with time according to
eq. [\ref{eq:responsomcr}].)

\subsection{Free Precession Encore}

\subsubsection{Different Crust and Superfluid Angular Velocities in Unperturbed State}
\label{sec:diffrot}

The fact that time-independent spindown of the crust implies a difference
between $\Omcrthree$ and $\Omsthree$ in steady state suggests that we
explore free precession once again, but with a different unperturbed state
than we used in Section \ref{sec:freeprec}. There, we assumed that the
undisturbed star rotates with $\Omvec_s=\Omvec_{cr}
=\Omega{\bf\hat e_3}$. Let us consider instead  what happens when
$\Omvec_c=\Omega{\bf\hat e_3}$ and $\Omvec_s=\xi\Omega{\bf\hat e_3}$ in the 
unperturbed state. We shall not assume that $\vert\xi-1\vert$ must be small,
although we expect this to be true.

The equations governing the precession of an axisymmetric star given this new
unperturbed state are
\be
\dot\Omcrtil-i\sigmapxi\Omcrtil+i{I_s\betap\over I_1}\Omstil
=-{I_s\beta\over I_1}(\Omcrtil-\Omstil)
\ee
\be
\dot\Omstil+i(1-\betap)(\Omstil-\xi\Omcrtil)=
-\beta(\Omstil-\Omcrtil),
\ee
where 
\be
\sigmapxi\equiv\epsilon+{\betap I_s\xi\over I_1}.
\ee
These equations yield the characteristic equation
\be
p^2+p\biggl[i(1-\betap-\sigmapxi)+\beta\biggl(1+{I_s\over I_1}\biggr)
\biggr]+\epsilon(1-\betap-i\beta)+i{\beta I_s\over I_1}(1-\xi)=0;
\ee
in the weakly coupled limit, the solutions are
\be
p_d=-i+{(i\betap-\beta)(1+\epsilon+\xi I_s/I_1)\over
		1+\epsilon}
\ee
\be
p_p=i\epsilon+{\{i\epsilon\betap\xi-\beta[\epsilon-(\xi-1)]\}I_s
\over I_1(1+\epsilon)},
\ee
and in the strong coupling limit,
\be
p_d=-{i\epsilon(1-\betap)+\beta[\epsilon+I_s(\xi-1)/I_1]
\over\epsilon+\xi I_s/I_1}
\ee
\be
p_p=i\sigmapxi-{I_s[i\xi(1-\betap)+\beta(1+\sigmapxi)]\over
I_1\sigmapxi}.
\ee
These equations are hardly different than their $\xi=1$ counterparts
except for two noteworthy differences. First, in the weakly coupled
limit, the real part of $p_p$ can become positive for $\xi-1>\epsilon$,
implying linear instability. However, this is not worrisome since 
the growth time of the mode is of order the spindown time of the star
if $\xi-1=N_{sd}/2\beta(I_s+I_3)$. 
Second, in the strongly coupled
limit, the real part of $p_d$ is enhanced for $\xi-1>0$, implying 
faster damping. Otherwise, the effect of $\xi\neq 1$ is merely to renormalize
the various coefficients appearing in the solutions to the characteristic
equation without introducing any qualitatively new behavior. Consequently,
we shall not consider the effects of differential rotation further in
this paper.

\subsubsection{Interaction of Two Regions with Simple Modes}\label{sec:simple}

In Sections \ref{sec:axi} and \ref{sec:nonaxi}, we explored the characteristic
modes of precession for axisymmetric and nonaxisymmetric crusts in the limits
of weak and strong coupling to the crustal superfluid. In general, we can
imagine that there are regions of the crust to which the superfluid couples
with different strengths. These regions are presumably linked to one
another by elastic forces that seek to enforce corotation.

To get a crude idea of what might transpire in such a situation, let us
imagine dividing the crust into two components, $a$ and $b$. To get a
schematic feeling for the normal modes of the coupled system, suppose
the angular velocities can be described by the equations
\be
\dot\Omvecatil-p_a\Omvecatil=-{\gamma I_b\over I_a+I_b}(\Omvecatil-\Omvecbtil)
\label{eq:omvecaeqn}
\ee
\be
\dot\Omvecbtil-p_b\Omvecbtil=-{\gamma I_a\over I_a+I_b}(\Omvecbtil-\Omvecatil),
\ee
where $\Omvecatil=\Omvec_{a,1}+i\Omvec_{b,1}$ and $\Omvecbtil=\Omvec_{b,1}
+i\Omvec_{b,2}$. This system of equations has the characteristic equation
\be
0=p^2-p(p_a+p_b-\gamma)+p_ap_b-{\gamma(p_aI_a+p_bI_b)\over I_a+I_b}.
\ee
When $\gamma$ is small, we rewrite the above equation as
\be
0=(p-p_a)(p-p_b)+\gamma\biggl(p-{p_aI_a+p_bI_b\over I_a+I_b}\biggr);
\ee
to first order in $\gamma$, the roots are
\be
p_1=p_a-{\gamma I_b\over I_a+I_b}
\ee
\be
p_2=p_b-{\gamma I_a\over I_a+I_b}.
\label{eq:compbwk}
\ee
The effect of weak coupling is additional damping of the modes of the
individual components. 
When $\gamma$ is large, we rewrite the characteristic
equation as
\be
0=p-{p_aI_a+p_bI_b\over I_a+I_b}
+\gamma^{-1}(p-p_a)(p-p_b).
\ee
In this limit, the two roots are
\be
p_1=-\gamma+{p_aI_b+p_bI_a\over I_a+I_b}
\ee
\be
p_2={p_aI_a+p_bI_b\over I_a+I_b}
	-{(p_b-p_a)(I_a-I_b)\over\gamma(I_a+I_b)}.
\ee
In this case, $p_1$ represents almost pure damping at a rate close to
$\gamma$, whereas $p_2$ represents damped precession at a rate that is
approximately the sum of the precession frequencies of the individual
components weighted by their moment of inertia fractions.

Consequently, if the characteristic coupling timescales among different
components of the crust are short, we find
a mean precession frequency weighted toward regions that comprise the
bulk of the crustal moment of inertia.
Only if the coupling times are long will the precession
frequencies of individual crustal components be apparent.
This indicates that even if there are small regions of unpinned superfluid
in the crust, precession at the Euler rate appropriate to those zones
may only be seen if they do not couple to the rest of the crust efficiently,
and that even if this is so, the precession will damp out eventually as
a consequence of the dissipative interaction. 

From time to time, it has also been suggested that Tkachenko modes of the
core superfluid could manifest themselves in the observed rotation of a
pulsar, which is presumably the angular velocity of its crust. If the
crustal superfluid is strongly pinned, we can regard one component,
$a$, as the crust, with $p_a=\sigma=\epsilon+I_s/I_1$, and the other as the
core, with $p_b$ the oscillation frequency of the core in the zero-coupling
limit. Equation (\ref{eq:compbwk}) shows that if the interaction between
crust and core is weak, there is indeed a mode of oscillation close to
$p_b$. For small but nonzero $\gamma$, the oscillations decay at a rate
$\gamma I_a/(I_a+I_b)$, which implies a damping time $\sim\gamma^{-1}I_c/I_1$
rotation periods, if we assume $I_a=I_1\ll I_b=I_c$; this would be of order
$(400-10^4)I_c/I_1$ for the damping times estimated by Alpar and Sauls (1988),
or approximately days to months for $I_c/I_1\sim 10^2$, comparable to  
the estimated periods of oscillation if they can occur. 
Moreover, for small
values of $\gamma$, the effect of the core oscillations on the crustal
angular velocity is small: equation (\ref{eq:omvecaeqn}) implies that
\be
\Omvecatil=\biggl[{\gamma I_b\over (I_b+I_a)(p_b-p_a)+\gamma I_b}\biggr]
\Omvecbtil,
\ee
which means that the amplitude of the oscillations in the angular velocity
of the crust is $\sim\gamma I_b/p_a(I_b+I_a)$ times the amplitude of
the oscillations of the angular velocity of the core, assuming that
$\vert p_a\vert\gg\vert p_b\vert$. Thus, even if the coupling is so
weak that the damping time $t_d$ associated with core oscillations is very
long, their observable manifestation in the angular velocity of the
crust is suppressed by a factor $\sim (\Omega t_d)^{-1}$.

\section{THREE COMPONENT STAR}

The calculations presented in Section \ref{sec:twocomp} are valid if the crust
and crustal superfluid are either completely decoupled from the core of the
star or coupled to it perfectly. In the limit of perfect pinning, Shaham
(1977) found that precession was damped as a consequence of imperfect
coupling to the core. Here, we examine a model consisting of three components, 
two of which are superfluid components that couple directly to the rigid
crust. We have in mind two possible applications, one in which the three
components are rigid crust, crustal superfluid and core (super)fluid, and
another in which the three components are rigid crust and two different
regions of crustal superfluid with different frictional couplings to the
rigid crust.

\subsection{Free Precession}\label{sec:frprec3}

\subsubsection{Crust, Crustal Superfluid and Core (Super)fluid}
\label{sec:crcscor}

We assume that the core couples directly only to the rigid component of the
crust, via a torque of the form
\be
\Nvec_{cc}=-I_c\zeta\Omega_c(\Omcore-\Omcr)+I_c\zetap(\Omcr\times\Omcore),
\ee
for small differences between the angular velocities of the crust and core,
which are both nearly $\Omega{\bf\hat e_3}$. This form of the torque is
analogous to equations (\ref{eq:nvecfinal}) and (\ref{eq:nvecpfinal}), except 
that the dissipative torque has been assumed to be isotropic (by contrast
to $\nvec$). We have also included a non-dissipative torque in $\Nvec_{cc}$,
unlike Shaham (1977; see Section \ref{sec:sh77}); this contribution can be
ignored by setting $\zetap=0$. Below, we shall assume that $\zeta\sim\zetap$,
at least for keeping track of small quantities.

If we define
\be
\Delomcor=\Omcorone-\Omcrone+i(\Omcortwo-\Omcrone),
\ee
then the coupled equations for the three component star may be written in the
form
\be
\dot\Omcrtil-i\epsilon\Omcrtil+{I_s\over I_1}(i\betap-\beta)\Delomtil
+{I_c\over I_1}(i\zetap-\zeta)\Delomcor=0
\label{eq:ccscorcr}
\ee
\be
\dot\Delomtil+
\biggl\{i\biggl[1-\betap\biggl(1+{I_s\over I_1}\biggr)\biggr]
+\beta\biggl(1+{I_s\over I_1}\biggr)\biggr\}\Delomtil
+i\epsilon\Omcrtil
-{I_c\over I_1}(i\zetap-\zeta)\Delomcor=0
\label{eq:ccscorcs}
\ee
\be
\dot\Delomcor+
\biggl\{i\biggl[1-\zetap\biggl(1+{I_c\over I_1}\biggr)\biggr]
+\zeta\biggl(1+{I_c\over I_1}\biggr)\biggr\}\Delomcor
+i\epsilon\Omcrtil
-{I_s\over I_1}(i\betap-\beta)\Delomtil=0.
\label{eq:ccscorcor}
\ee
Assuming modes proportional to $\exp(p\phi)$, we find that
\be
\Delomtil=-{p\Omcrtil\over p+i(1-\betap)+\beta}
\qquad{\rm and}\qquad
\Delomcor=-{p\Omcrtil\over p+i(1-\zetap)+\zeta}.
\label{eq:modes3}
\ee
The third order characteristic equation for this system may be
written in the form
\begin{eqnarray}
0 & = & (p+i)\biggl\{p^2
+p\biggl[i(1-\betap-\sigmap)+\beta\biggl(1+{I_s\over I_1}\biggr)\biggr]
+\epsilon(1-\betap-i\beta)\biggr\}
\nonumber\\ & &
-(i\zetap-\zeta)\biggl[p^2\biggl(1+{I_c\over I_1}\biggr)
+p\biggl\{i\biggl[(1-\betap)\biggl(1+{I_c\over I_1}\biggr)-\sigmap\biggr]
+\beta\biggl(1+{I_s+I_c\over I_1}\biggr)\biggr\}\nonumber\\ & &
+\epsilon(1-\betap-i\beta)\biggr],
\label{eq:char3}
\end{eqnarray}
where $\sigmap$ is defined in equation (\ref{eq:sigmapdef}).

When $\zeta$ and $\zetap$ are small, two of the solutions of equation
(\ref{eq:char3}) are close to the solutions $p_d$ and $p_p$ of the 
second order equation (\ref{eq:axichar});
to first order in $\zeta$ and $\zetap$, the corrections to $p_d$ and
$p_p$ are
\be
\delta p_d={(I_c/I_1)(i\zetap-\zeta)
[p_d(i\sigmap-\beta I_s/I_1)-\epsilon(1-\betap-i\beta)]
\over (i+p_d)(p_d-p_p)}
\label{eq:dpdzwk}
\ee
\be
\delta p_p={(I_c/I_1)(i\zetap-\zeta)
[p_p(i\sigmap-\beta I_s/I_1)-\epsilon(1-\betap-i\beta)]
\over (i+p_p)(p_p-p_d)},
\label{eq:dppzwk}
\ee
respectively. The third solution is a new mode near $p=-i$; to first
order in small quantities it is
\be
p_d^\prime=-i+{i(i\betap-\beta)(i\zetap-\zeta)
[1+\epsilon+(I_s+I_c)/I_1]\over
(i+p_d)(i+p_p)}.
\ee
When $\beta$ and $\betap$ are small, we use equations (\ref{eq:pdbwk})
and (\ref{eq:axiprecweak}) in equations (\ref{eq:dpdzwk}) and (\ref{eq:dppzwk})
to find
\be
\delta p_d={I_c(i\zetap-\zeta)(\epsilon+I_s/I_1)\over I_1(1+\epsilon)}
\ee
\be
\delta p_p={I_c\epsilon(i\zetap-\zeta)\over I_1(1+\epsilon)};
\label{eq:ppbwkzwk}
\ee
the new mode is
\be
p_d^\prime=-i+{(i\zetap-\zeta)(1+\epsilon+I_s/I_1+I_c/I_1)\over
1+\epsilon+I_s/I_1}.
\ee
When $\beta$ and $1-\betap$ are small, upon using equations (\ref{eq:pdbstr}) 
and (\ref{eq:axiprecstrong}) in equations (\ref{eq:dpdzwk}) and (\ref{eq:dppzwk}) 
we find that $\delta p_d$ is higher order in $\beta$ and $1-\betap$, and hence 
very small, while 
\be 
\delta p_p={I_c(i\zetap-\zeta)\sigmap\over I_1(1+\sigmap)} 
\label{eq:ppbstrzwk}
\ee 
\be 
p_d^\prime=-i+{(i\zetap-\zeta)(1+\sigmap+I_c/I_1)\over 1+\sigmap}; 
\label{eq:pdpbstrzwk}
\ee 
these two results are equivalent to equations (\ref{eq:ppshwk})
and (\ref{eq:pdshwk}) of Section \ref{sec:sh77} if we substitute
$\sigma$ for $\sigmap$ and $-\gamma$ for $(i\zetap-\zeta)(1+I_c/I_1)$.
The qualitative conclusion reached by Shaham (1977) for weak crust-core
coupling is duplicated here: according to equations (\ref{eq:ppbwkzwk})
and (\ref{eq:ppbstrzwk}), precession damps out in $\sim I_1/I_c\zeta$
precession periods irrespective of the effectiveness of vortex drag.
We note, though, that the mode corresponding to $p_d^\prime$ implies
angular velocities that are nearly but not precisely fixed in the
inertial frame of the observer; if $\zetap>\zeta$, these could complete
at least one period of oscillation before decaying away (although we do
not expect this to be the case generally). Such modes are
also found in studies of the rotation of the Earth, where they may arise
from departures from rigid rotation of the fluid core confined by
the overlying crust; for the Earth, the result is a retrograde motion
of the pole (see Lambeck 1980, Section 3.3 for a physical and historical
review). Since the mode arises because the core angular velocity 
remains fixed in the inertial frame when $\zeta$ and $\zetap$ are
zero identically, we expect that for small but finite crust-core
coupling, the mode corresponds principally to oscillations of the
angular velocity of the core. From equation (\ref{eq:modes3}) we
find
\be
{\Omcrtil\over\Delomcor}\approx(\zetap+i\zeta)
\biggl(\sigmap+{I_c\over I_1}\biggr)\approx
(\zetap+i\zeta)\biggl({I_s+I_c\over I_1}\biggr)
\label{eq:modewander}
\ee
for this mode, which decreases linearly with the frequency of oscillation,
but may be substantial nevertheless if $I_s+I_c\gg I_1$.

For large values of $\vert i\zetap-\zeta\vert$, one of the roots of
equation (\ref{eq:char3}) is
\be
p_d^\prime=(i\zetap-\zeta)\biggl(1+{I_c \over I_1}\biggr)
-i\biggl(1-{\sigmap I_c\over I_c+I_1}\biggr),
\ee
which is equivalent to equation (\ref{eq:pdshstr}) if we substitute
$\sigma$ for $\sigmap$ and 
$-\gamma$ for $(i\zetap-\zeta)(1+I_c/I_1)$; this root does not depend
on the strength of the vortex drag to lowest order in 
$\vert i\zetap-\zeta\vert^{-1}$. When vortex drag is weak, the other two
roots of equation (\ref{eq:char3}) are
\be
p_d=-i+{(i\betap-\beta)(1+\epsilon+I_s/I_1+I_c/I_1)\over
1+\epsilon+I_c/I_1}
\label{eq:pdbwkzstr}
\ee
\be
p_p={i\epsilon\over 1+I_c/I_1}
+{I_s\epsilon(i\betap-\beta)\over I_1(1+\epsilon+I_c/I_1)}
-{I_c\epsilon(i\zetap+\zeta)(1+\epsilon+I_c/I_1)\over
I_1[(\zetap)^2+\zeta^2](1+I_c/I_1)^3}.
\ee
(Corrections to eq. [\ref{eq:pdbwkzstr}] proportional to 
$\vert i\zetap-\zeta\vert^{-1}$ are higher order in $\beta$
and $\betap$, and have been dropped.) In the limit of strong
vortex drag,
\be
p_d=-{\epsilon[i(1-\betap)+\beta]\over\sigmap}
\label{eq:pdbstrzstr}
\ee
\be
p_p={i\sigmap\over 1+I_c/I_1}
-{I_s\over I_1\sigmap}\biggl\{i(1-\betap)+\beta
\biggl[1+\sigmap-{I_sI_c\over I_1(I_1+I_c)}\biggr]
\biggr\}
-{I_c\sigmap(i\zetap+\zeta)(1+\epsilon+I_c/I_1)\over
I_1[(\zetap)^2+\zeta^2](1+I_c/I_1)^3}.
\label{eq:ppbstrzstr}
\ee
(Corrections to eq. [\ref{eq:pdbstrzstr}] proportional to
$\vert i\zetap-\zeta\vert^{-1}$ are higher order in $\beta$
and $1-\betap$, and have been dropped.) The correction for
strong but imperfect crust-core coupling in equation 
(\ref{eq:ppbstrzstr}) is equivalent to equation (\ref{eq:ppshstr})
if $-\gamma$ is substituted for $(i\zetap-\zeta)(1+I_c/I_1)$.

Interaction with the core of the star enhances the damping of the
precessing modes in all of the limiting cases explored above.
For very small or very large coupling between the crust and
crustal superfluid, the crust-core interactions are the principal
cause of decay. In the strongly pinned regime, the characteristic
timescales for decay are just what Shaham (1977) estimated. Imperfect
pinning allows a new eigenvalue $p_d$, but the associated mode
damps out quickly, and so cannot be the explanation for observations
of persistent cyclical variations in pulsar spin rates.

\subsubsection{Crust Coupled to Two Different Regions of Crustal Superfluid}
\label{sec:crtwocs}

The equations governing the spin dynamics of this system are the same
as equations (\ref{eq:ccscorcr}), (\ref{eq:ccscorcs}) and (\ref{eq:ccscorcor})
if we identify $\Omcore$ with the angular velocity of the second crustal
superfluid component, and $\zeta$ and $\zetap$ with the coefficients coupling
this component to the rigid crust. Then it is clear that the characteristic
equation for the normal modes of this system is still equation (\ref{eq:char3}),
which we rewrite in the form
\begin{eqnarray}
0 & = & p^3-ip^2\biggl(\epsilon+{I_s+I_c\over I_1}\biggr)
\nonumber\\ & &
+(1-\betap-i\beta)\biggl[ip^2\biggl(1+{I_s\over I_1}\biggr)
+p\biggl(\epsilon+{I_c\over I_1}\biggr)\biggr]\nonumber\\ & &
+(1-\zetap-i\zeta)\biggl[ip^2\biggl(1+{I_c\over I_1}\biggr)
+p\biggl(\epsilon+{I_s\over I_1}\biggr)\biggr]\nonumber\\ & &
+(1-\betap-i\beta)(1-\zetap-i\zeta)
\biggl[i\epsilon-p\biggl(1+{I_s+I_c\over I_1}\biggr)\biggr],
\end{eqnarray}
which exhibits symmetry under interchange of superfluid components 
explicitly.

This form of the characteristic equation is especially useful when 
both superfluid components couple strongly to the rigid crust. In
that limit, one of the roots is
\be
p_p=i\sigmapp-\biggl({1+\sigmapp\over\sigmapp}\biggr)
\biggl\{[i(1-\betap)+\beta]{I_s\over I_1}
	   +[i(1-\zetap)+\zeta]{I_c\over I_1}\biggr\}
\ee
to first order in small quantities, where
\be
\sigmapp\equiv\epsilon+{I_c+I_s\over I_1};
\ee
the appearance of
this root suggests a simple generalization to a multicomponent
superfluid, with separate moments of inertia $I_{s,j}$ and coupling
coefficients $\beta_j$ and $\betap_j$:
\be
p_p=i\sigmapp
-\biggl({1+\sigmapp\over\sigmapp}\biggr)
\sum_j[i(1-\betap_j)+\beta_j]{I_{s,j}\over I_1},
\ee
where 
\be
\sigmapp\equiv\epsilon+\sum_j{I_{s,j}\over I_1}.
\ee
The other two roots are first
order small to leading order; they are approximately equal to the
two roots of the quadratic equation
\begin{eqnarray}
0 & = & p^2\sigmapp+
ip\biggl[(1-\betap-i\beta)\biggl(\epsilon+{I_c\over I_1}\biggr)
\nonumber \\ & &
+(1-\zetap-i\zeta)\biggl(\epsilon+{I_s\over I_1}\biggr)\biggr]
-\epsilon(1-\betap-i\beta)(1-\zetap-i\zeta).
\end{eqnarray}
When the crust is only slightly nonspherical, so $\epsilon\ll 1$,
the two roots of this equation are approximately
\be
p_+\approx -{[\Bvar](I_c/I_1)+[\Zvar](I_s/I_1)\over\sigmapp}
\ee
\be
p_-\approx -{\epsilon[\Bvar][\Zvar]\over
[\Bvar](I_c/I_1)+[\Zvar](I_s/I_1)}.
\ee
Notice that when one of the superfluid components is 
coupled to the rigid crust more strongly
than the other, $p_+$ is dominated by the
tighter-coupled component, but the slow mode $p_-$ is dominated
by the weaker-coupled one, and reduces to equation
(\ref{eq:pdbstr}), with $\sigmap\approx I_s/I_1$ if 
$I_s\vert\Zvar\vert\gg I_c\vert\Bvar\vert$ and $\sigmap\approx I_c/I_1$
if $I_c\vert\Bvar\vert\gg I_s\vert\Zvar\vert$.

The limit in which one component couples strongly to the crust while
the other couples only weakly is also of interest potentially, 
particularly in the aftermath of a pulsar glitch, in which some parts
of the crustal superfluid may decouple rapidly and recouple only
slowly if at all (e.g. Sedrakian 1995). The modes for that situation
are given by equation (\ref{eq:axiprecstrong}) with the correction
given in equation (\ref{eq:ppbstrzwk}), equation (\ref{eq:pdbstr}) and
equation (\ref{eq:pdpbstrzwk}). If $\zetap>\zeta$, it is possible that
the mode corresponding to $p_d^\prime$ given by equation (\ref{eq:pdpbstrzwk})
yields slowly damped oscillations in the angular velocity, as was discussed
in Section \ref{sec:crcscor}.

\subsection{Response to External Torques}

This section is analogous to Section \ref{sec:torq2}, except that 
we need to consider three distinct external torques, acting on 
the crust, crustal superfluid and core, respectively, and the
linear response of the three different components to each.

\subsubsection{Response to External Torques on the Crust}

When the crust is subject to an external torque $\Nvec_{cr}(\phi)$, 
\be
\dot\Omcrtil-i\epsilon\Omcrtil+{I_s\over I_1}(i\betap-\beta)\Delomtil
+{I_c\over I_1}(i\zetap-\zeta)\Delomcor=\Nveccr(\phi)
\ee
\be
\dot\Delomtil+
\biggl\{i\biggl[1-\betap\biggl(1+{I_s\over I_1}\biggr)\biggr]
+\beta\biggl(1+{I_s\over I_1}\biggr)\biggr\}\Delomtil
+i\epsilon\Omcrtil
-{I_c\over I_1}(i\zetap-\zeta)\Delomcor=-\Nveccr(\phi)
\ee
\be
\dot\Delomcor+
\biggl\{i\biggl[1-\zetap\biggl(1+{I_c\over I_1}\biggr)\biggr]
+\zeta\biggl(1+{I_c\over I_1}\biggr)\biggr\}\Delomcor
+i\epsilon\Omcrtil
-{I_s\over I_1}(i\betap-\beta)\Delomtil=-\Nveccr(\phi).
\ee
It is straightforward to show that these equations have the
particular solution
\be
\Omcrtil=\sum_{p,d,d^\prime}A_\alpha
\int_{-\infty}^\phi{d\phip\Nveccr(\phip)
\exp[p_\alpha(\phi-\phip)]}
\label{eq:omcrcr3}
\ee
\be
\Delomtil=-\sum_{p,d,d^\prime}{p_\alpha A_\alpha\over
p_\alpha+\Bvar}
\int_{-\infty}^\phi{d\phip\Nveccr(\phip) 
\exp[p_\alpha(\phi-\phip)]} 
\ee
\be
\Delomcor=-\sum_{p,d,d^\prime}{p_\alpha A_\alpha\over
p_\alpha+\Zvar}
\int_{-\infty}^\phi{d\phip\Nveccr(\phip)  
\exp[p_\alpha(\phi-\phip)]},
\ee
where the coefficients are
\be
A_p={[p_p+\Bvar][p_p+\Zvar]\over (p_p-p_d)(p_p-p_d^\prime)}
\ee
\be
A_d={[p_d+\Bvar][p_d+\Zvar]\over (p_d-p_d^\prime)(p_d-p_p)}
\ee
\be
A_d^\prime={[p_d^\prime+\Bvar][p_d^\prime+\Zvar]\over
(p_d^\prime-p_d)(p_d^\prime-p_p)}.
\ee
Qualitatively, these are similar to what we found for the response of a two
component star, except that the response timescales are shorter as a consequence
of crust-core coupling, which implies (perhaps significantly)
enhanced decay of the modes. Notice that as the coupling between the rigid
crust and crustal superfluid becomes perfect, $\Delomtil\to -\Omcrtil$, and
the response simplifies to
\begin{eqnarray}
\Omcrtil & = &(p_p-p_d^\prime)^{-1}\int_{-\infty}^\phi{d\phip\Nveccr(\phip)
\{[p_p+\Zvar]\exp[p_p(\phi-\phip)]}\nonumber\\ & & \qquad\qquad\qquad
-[p_d^\prime+\Zvar]\exp[p_d^\prime
(\phi-\phip)]\}
\label{eq:omcr3bs}
\end{eqnarray}
\be
\Delomcor=(p_p-p_d^\prime)^{-1}\int_{-\infty}^\phi{d\phip\Nveccr(\phip)
\{-p_p\exp[p_p(\phi-\phip)]+p_d^\prime\exp[p_d^\prime(\phi-\phip)]\}};
\ee
these results are identical with equations (\ref{eq:responsomcr}) and (\ref{eq:responsedelom})
for the response of a two component star, with the substitution of
$\Delomcor$ for $\Delomtil$, $p_d^\prime$ for $p_d$, and $\Zvar$ for $\Bvar$.
However, their usefulness is restricted to $\Nveccr(\phi)$ that vary on timescales short compared
with the characteristic damping time implied by $p_d$, the slowest decaying mode,
as was discussed in Section \ref{sec:torq2}.

The response to a time independent torque on the crust is simply $\Omcrtil=i\Nveccr/\epsilon$,
just as in the two component case, with $\Delomtil=\Delomcor=0$. These are the approximate
responses obtained for a constant torque originating at a finite time $\phi_0$ in the past
provided that $p_d(\phi-\phi_0)\gg 1$. 

\subsubsection{Response to Torques on the Core (Super)Fluid}

If the core is subject to a torque $\Nvec_c(\phi)$ then
\be
\dot\Omcrtil-i\epsilon\Omcrtil+{I_s\over I_1}(i\betap-\beta)\Delomtil
+{I_c\over I_1}(i\zetap-\zeta)\Delomcor=0
\ee
\be
\dot\Delomtil+
\biggl\{i\biggl[1-\betap\biggl(1+{I_s\over I_1}\biggr)\biggr]
+\beta\biggl(1+{I_s\over I_1}\biggr)\biggr\}\Delomtil
+i\epsilon\Omcrtil
-{I_c\over I_1}(i\zetap-\zeta)\Delomcor=0
\ee
\be
\dot\Delomcor+
\biggl\{i\biggl[1-\zetap\biggl(1+{I_c\over I_1}\biggr)\biggr]
+\zeta\biggl(1+{I_c\over I_1}\biggr)\biggr\}\Delomcor
+i\epsilon\Omcrtil
-{I_s\over I_1}(i\betap-\beta)\Delomtil=\Nveccor(\phi),
\ee
where $\Nveccor\equiv I_c^{-1}(\Nvec_{c,1}+i\Nvec_{c,2})$.
Apart from decaying transients, the solution to these equations is
\be
\Omcrtil=\sum_{p,d,d^\prime}B_\alpha
\int_{-\infty}^\phi{d\phip\Nveccor(\phip)
\exp[p_\alpha(\phi-\phip)]}
\label{eq:t3scr}
\ee
\be
\Delomtil=-\sum_{p,d,d^\prime}{p_\alpha B_\alpha\over
p_\alpha+\Bvar}
\int_{-\infty}^\phi{d\phip\Nveccor(\phip)
\exp[p_\alpha(\phi-\phip)]}
\label{eq:t3ss}
\ee
\be
\Delomcor=-\sum_{p,d,d^\prime}{p_\alpha B_\alpha\over
p_\alpha+\Zvar}
\int_{-\infty}^\phi{d\phip\Nveccor(\phip) 
\exp[p_\alpha(\phi-\phip)]},
\label{eq:t3scor}
\ee
where the coefficients are
\be
B_p={[p_p+\Zvar][p_d+\Zvar][p_d^\prime+\Zvar][p_p+\Bvar]\over
[\Zvar][\BmZvar](p_p-p_d)(p_p-p_d^\prime)}
\ee
\be
B_d={[p_p+\Zvar][p_d+\Zvar][p_d^\prime+\Zvar][p_d+\Bvar]\over
[\Zvar][\BmZvar](p_d-p_p)(p_d-p_d^\prime)}
\ee
\be
B_d^\prime={[p_p+\Zvar][p_d+\Zvar][p_d^\prime+\Zvar][p_d^\prime+\Bvar]\over
[\Zvar][\BmZvar](p_d^\prime-p_p)(p_d^\prime-p_d)}.
\ee

When the crust and crustal superfluid
are coupled to one another perfectly, $p_d/[\Bvar]\to -\epsilon/\sigmap$,
and $B_d\to 0$, so $\Delomtil\to-\Omcrtil$ and
\begin{eqnarray}
\Omcrtil & = &-{[p_p+\Zvar][p_d^\prime+\Zvar]\over[\Zvar](p_p-p_d^\prime)}
\int_{-\infty}^\phi{d\phip\Nveccor(\phip)\{\exp[p_p(\phi-\phip)]}
\nonumber\\ & & \qquad\qquad\qquad\qquad\qquad\qquad\qquad
\qquad\qquad\qquad
-\exp[p_d^\prime(\phi-\phip)]\}
\end{eqnarray}
\begin{eqnarray}
\Delomcor &=&\int_{-\infty}^\phi{d\phip\Nveccor(\phip)
\biggl\{{p_p[p_d^\prime+\Zvar]\exp[p_p(\phi-\phip)]
\over[\Zvar](p_p-p_d^\prime)}}\nonumber\\ & &
\qquad\qquad\qquad\qquad
-{p_d^\prime[p_p+\Zvar]\exp[p_d^\prime(\phi-\phip)]
\over[\Zvar](p_p-p_d^\prime)}\biggr\};
\end{eqnarray}
As was discussed in Section \ref{sec:torq2},
this simplified version of the response
may only be used for torques that vary rapidly compared to the 
damping timescale implied by $p_d$.

\section{CONCLUSIONS}

Shaham (1977) demonstrated that when superfluid pins perfectly
to crustal nuclei, the precession period of a neutron star is
shortened immensely, and, moreover, the precession damps quickly
as a result of weak coupling to the stellar core. One of the
principal goals of this paper has been to examine whether there
are additional modes with long periods and long damping timescales
when the assumption of perfect coupling between crustal nuclei
and superfluid is relaxed. In fact, when the coupling is strong
but imperfect, there are new modes that have
very long characteristic timescales.
One new mode is given by equation (\ref{eq:pdbstr}),
$$
p_d=-{\epsilon[\beta+i(1-\betap)]
\over\sigmap}
\eqno(\ref{eq:pdbstr})
$$
for an axisymmetric, two component star, where $\epsilon$ is 
the oblateness of the star and $\sigmap=\epsilon+I_s/I_1$,
where $I_s$ is the moment of inertia of the crustal superfluid
and $I_1$ one of the principal moments of inertia of the crust.
(Coupling of the crust to the stellar core hardly alters this
result; see discussion in Section \ref{sec:frprec3}.)
Since $\beta\ll 1$, in the limit of strong vortex drag, this
mode is extremely long-lived; moreover, since $1-\betap\ll 1$
in this limit, the mode undergoes oscillations that are also
extremely slow. The problem is that we expect that $\vert 1-\betap\vert
\sim\beta^2\ll\beta$ in the strong coupling domain as long as the
vortex drag coefficient $\etap$, which governs the strength of the
drag force perpendicular to the direction of motion of the vortex through
the normal fluid, is small compared with $\eta$, the analogous coefficient
for the strength of the drag force antiparallel to the direction of motion
(e.g. eqs.
[\ref{eq:betadef}] and [\ref{eq:betapdef}] and ensuing discussion). 
Thus, this mode is not actually oscillatory at all, for it damps
before it can complete a single cycle. In fact, for a nonaxisymmetric
star, $p_d$ splits into two modes, with (see eq. [\ref{eq:pdnonaxi}])
$$
p_d^\pm=-{\beta(\epsone\sigmap_2+\epstwo\sigmap_1)\over
2\sigmap_1\sigmap_2}
\pm
\biggl[{\beta^2(\epsone\sigmap_2-\epstwo\sigmap_1)^2\over
(2\sigmap_1\sigmap_2)^2}
-{\epsone\epstwo(1-\betap)^2\over\sigmap_1\sigmap_2}
\biggr]^{1/2},
\eqno(\ref{eq:pdnonaxi})
$$
both of which may be purely real and decaying. 

Another new mode arises in three component models when, for example,
the crustal
superfluid is coupled strongly to the rigid crust in some regions
and weakly in others, or else the crustal superfluid is strongly
coupled to the rigid crust but the superfluid core is coupled to
it only weakly. Under such circumstances, one solution to the
three component characteristic equation is equation 
(\ref{eq:pdpbstrzwk})
$$
p_d^\prime=-i+{(i\zetap-\zeta)(1+\sigmap+I_c/I_1)\over 1+\sigmap}
\eqno(\ref{eq:pdpbstrzwk})
$$
where $\zetap$ and $\zeta$ are the coupling parameters between the
rigid crust and the component that is barely tied to it. As was 
discussed in Section \ref{sec:crcscor}, this mode can lead to
a slow wandering of the pole of the neutron star as seen in the
inertial reference frame. However, the excitation amplitude is
relatively small for the crustal angular velocity in this mode
(e.g. eq. [\ref{eq:modewander}]); moreover, in the weak 
coupling domain, we expect $\zetap\ll\zeta$ if $\etap\ll\eta$
(e.g. Section \ref{sec:setup}), so the mode decays before
completing one oscillation. 

Thus, it appears likely that although there are new, possibly
long-lived modes for a neutron star with strong but imperfect coupling
between superfluid and rigid crust, these modes are not
principally oscillatory as long as $\etap\ll\eta$. Only if 
there are regions in the star
where this inequality is reversed somehow
could damped oscillations occur.

There may be regions of weak coupling between crustal superfluid
and nuclei interspersed among regions of strong coupling. If so
these regions could, if tied to the strong coupling regions
tenously, undergo nearly independent oscillations with both
long cycle times and insignificant damping. Moreover, there could
be regions of the core that may undergo long period oscillations
if detached from the crust. However, in both cases, the effect of
nearly independent, slow and persistent oscillations on the
portion of the crust where superfluid vortex lines are
pinned would be minimal, tending to zero in the limit of
complete independence. Thus, if slow, persistent oscillations
can occur {\it somewhere} in the star, the chances that one can
know about them from observations of the rotation rate of that
part of the crust where superfluid is pinned strongly are remote.
In the opposite limit, in which all regions of the star are
coupled to one another closely, the observed frequencies
are averages weighted by moment of inertia, and tend to be
dominated by regions of high frequency and/or large moment of
inertia.

Under the combined action of external and internal torques,
the angular velocity of the crust tends to tilt away from
alignment with its principal axes. If the external torque
is time-independent, or only varies on a very long timescale,
then {\it ultimately}
the tilt angle approaches a constant value $\theta\sim
\vert N_{ex}\vert/\epsilon I_{cr}\Omega^2$, where $N_{ex}$ is the value of
the constant external torque, $I_{cr}$ is the typical
moment of inertia of the crust, and $\epsilon$ is the crustal
oblateness. If $N_{ex}=I\dot\Omega$ is the spindown torque,
where $I$ is the moment of inertia of the star, then the
steady state
tilt angle is $\sim -(I/\epsilon I_{cr})\dot\Omega/\Omega^2$,
where $I$ is the total moment of inertia of the star. Even
though $-\dot\Omega/\Omega^2\sim (\Omega t_{ds})^{-1}$, where
$t_{sd}$ is the spindown timescale, is very small ($\approx
5\times 10^{-9}P({\rm s})/t_{sd}({\rm y})$)
$I/\epsilon I_{cr}$ may be very large, and $\theta$ could
be non-negligible. An amusing side-effect of this tilt is
that even an axisymmetric neutron star could be a source
of gravitational radiation, with an amplitude that can be
determined from observables (e.g. spindown timescale,
period), quantities that can be inferred observationally
with varying degrees of confidence (e.g. distance) and
theoretically determined parameters (e.g. total moment of
inertia) but does not depend on the oblateness $\epsilon$.
Unfortunately, the implied wave amplitudes (strain amplitude
$h\simless 10^{-30}$) are well below
the projected capabilities of the {\it advanced} LIGO.

One key assumption behind this estimate of the steady state
tilt angle is contained in the italicized word {\it ultimately} 
of the previous paragraph.
As was discussed in Section \ref{sec:torq2}, the asymptotic
value of $\theta$ is only attained if the slowest damped
mode of the star decays in a time short compared to
the timescales on which the external torque varies. Practically
speaking, this means that if the crustal nuclei and superfluid
are closely pinned, then the steady state tilt is 
approached on the damping time of $p_d$ (given by 
eq. [\ref{eq:pdbstr}]); if this is short compared with the
spindown timescale, then the asymptotic value of $\theta$
is reached. Otherwise, the tilt could be smaller --
somewhere between $\sim -(I/\sigmap I_{cr})\dot\Omega/\Omega^2$
and $-(I/\epsilon I_{cr})\dot\Omega/\Omega^2$ -- and time
variable. As was pointed out in Section \ref{sec:torq2},
the correct steady state tilt angle, and the evolution toward
that angle, could {\it not} be found from Shaham (1977), where
perfect coupling between crust and crustal superfluid was
assumed. The crux of the solution is in the timescales implied
by the imperfection of the pinning.

We began this paper by proposing to examine perturbations about
a particular fixed point of the equations governing the rotational
dynamics of a neutron star, the one corresponding to equal
angular velocities of all components lined up along one of the
principal axes of the crust (the one with largest moment of
inertia eigenvalue). We have only wavered from this program
briefly, in Section \ref{sec:diffrot}, where we considered
perturbations about a state in which the rigid crust and crustal
superfluid have parallel angular velocities with slightly different
magnitudes. However, in spite of 
the constancy of our approach, we have uncovered some hints that
it may be unrealistic, for when external torques are
taken into account, the correct fixed points may involve
tilts away from principal axes. In a sequel to this paper,
we shall investigate the implications of time dependent and
time independent tilts due to external torques, as well as
to internal torques we have neglected here.  

\bigskip
\noindent This research was supported in part by NSF Grants
AST-93-15375 and AST-95-30397, and
NASA Grant NAG 5-3097. A.S. gratefully 
acknowledges the support of the Max Kade Foundation.

\end{document}